\documentclass[12pt]{article}
\usepackage{amssymb,euscript}
\usepackage{amsmath}
\usepackage{multirow}

\numberwithin{equation}{section}
\def\be{\begin{equation}}
\def\ee{\end{equation}}
\def\bea{\begin{eqnarray}}
\def\eea{\end{eqnarray}}
\newcommand{\nn}{\nonumber}

\newcommand\R{\mathbb{R}}
\newcommand\Z{\mathbb{Z}}

\newcommand\diff{\mathrm{d}}
\newcommand{\ex}{\mathrm{e}}
\newcommand{\de}{\partial}

\newcommand{\frakg}{\mathfrak{g}}

\newcommand{\lie}{\pounds}
\newcommand{\sch}{Schr\"odinger}
\newcommand{\cO}{\mathcal{O}}

\def\ket#1{|#1\rangle}
\def\bra#1{\langle#1|}
\def\pairing#1#2{\left(#1,#2\right)}

\begin{document}
%%%%%%%%%%%%%%%%%%%%%%%%%%%%%%%%%%%%%%%%%%%%%%%%%%%%%%%%%%%%%%%%%%%%%%%%
\begin{titlepage}
\begin{center}
\today

\vskip 2.8 cm 
{\large \bf Comments on Galilean conformal\\[2.6mm]
field theories and their geometric realization}

\vskip 1.5cm
Dario Martelli$^1$ and Yuji Tachikawa$^2$
\vskip 0.8 cm

{$^1$ \em Department of Physics,\\
 Swansea University,\\
Singleton Park, Swansea, SA2 8PP, U.K.\\ }

\vskip 0.6 cm
and 
\vskip 0.6 cm

{$^2$ \em School of Natural Sciences,  \\
Institute for Advanced Study,\\
Einstein Drive, Princeton, New Jersey, U.S.A.}\\

\end{center}

\vskip 2cm

\begin{abstract}
\noindent
We discuss non-relativistic conformal algebras generalizing the \sch\ algebra.
One instance of these algebras is a conformal, acceleration-extended, Galilei algebra, which arises 
also as a contraction of the relativistic conformal algebra. In two dimensions, this admits an ``exotic'' central extension, whereby boosts do not commute. We study general properties of non-relativistic conformal field theories with such symmetry. 
We realize geometrically the symmetry in terms of a metric invariant under the exotic conformal Galilei algebra, although its signature is neither Lorentzian nor Euclidean. We comment on holographic-type calculations in this background. 
\end{abstract}

\end{titlepage}
\pagestyle{plain}
\setcounter{page}{1}
\newcounter{bean}
\baselineskip18pt

%%%%%%%%%%%%%%%%%%%%%%%%%%%%%%%%%%%%%%%%%%%%%%%%%%%%%%%%%%%%%%%%%%%%%

\tableofcontents

\section{Introduction}
\label{introduction}

Recently there has been considerable activity in extending the 
Anti-de Sitter/conformal field theory (AdS/CFT) correspondence 
to various low-energy physical systems. In many instances these systems are \emph{non-relativistic},
implying that their fundamental symmetry is  Galilean. 
One example of Galilean symmetry  is the \sch\ symmetry, which is the symmetry of the free \sch\ equation.
It was proposed in \cite{son,mcgreevy}  that some gravity backgrounds with \sch\ symmetry should be holographically dual to certain conformal quantum mechanics relevant for the description of cold atoms \cite{Nishida:2007pj}.

In \cite{Maldacena:2008wh,Herzog:2008wg,Adams:2008wt}
it was shown that the  backgrounds of \cite{son,mcgreevy}
arise as solutions of String Theory. 
One way to embed these geometries in String Theory 
is based on the fact that the field theories arise from the 
discrete light cone quantization (DLCQ) of  relativistic conformal field theories. 
At the level of the symmetry, it utilizes the fact 
that the \sch\ group is a subgroup of the relativistic conformal 
group \cite{BurdetPerrinSorba,Duval:1984cj,Duval:1990hj}.
On the gravity side this corresponds to the light-like compactification of AdS$_5$, or more generally to 
a ``twisted'' version of this, obtained by a chain of String dualities. On the field theory side, it
corresponds to (the DLCQ of) certain dipole deformations of the parent relativistic CFT \cite{Bergman:2000cw}. 
 Subtleties involved in these constructions were discussed in \cite{Maldacena:2008wh}.

A natural question to ask is whether there exist non-relativistic conformal field theories different from 
those enjoying \sch\ symmetry, that may be realized holographically as in \cite{son,mcgreevy}. 
There is an abundance of symmetry groups, comprising the Galilei 
group as a subgroup\footnote{Note that the Lifshitz group considered in \cite{Kachru:2008yh} is not of this type.}, 
and some of them are known to be symmetries of certain condensed matter 
problems. If one in addition demands conformal invariance, it turns out that there is a class
of conformal Galilei algebras, generalizing precisely the \sch\ algebra \cite{negro1,negro2}.  The non-relativistic 
Navier-Stokes equations are an example where these symmetries are relevant \cite{Bhattacharyya:2008kq,Fouxon:2008ik}. For more examples in condensed matter
see \cite{Horvathy:2006gi} and references therein.

An interesting conformal Galilei algebra, which will be the main focus of the paper, is closely related to the \sch\ one. The main differences with respect to the latter are that dilatations act as in the relativistic case,
$D: (t, \vec{x}) \to (\lambda t, \lambda \vec{x})$, and 
it contains new generators corresponding to constant
\emph{accelerations} \cite{Stichel:2003kh,Lukierski:2007nh}. 
Furthermore, this algebra does not admit the usual Galilean central extension by the mass operator $M$. It thus necessarily describes massless particles. On the other hand,  in $d=2$ spatial dimensions, it admits a different central extension, by an operator we denote $\Theta$. Following the literature, we will refer to this centrally extended conformal Galilei algebra as ``exotic'' \cite{Duval:2000xr}.
One motivation for studying the centrally extended algebra is that the central extension played a key role in the geometric realization of the \sch\ symmetry, through the Bargmann construction \cite{Duval:1984cj,Duval:1990hj,Duval:2008jg}. This is related to the fact that in non-relativistic quantum mechanics wave functions
transform ``up to a phase'' \cite{Bargmann:1954gh} under the symmetry group, while they form true representations for the centrally extended group. 
Then  one would expect that adding an extra direction, conjugate to $\Theta$,  the
exotic conformal Galilei  algebra should be realized geometrically. In fact, this is roughly correct, as we will show. 

Another intriguing feature of this algebra is that it arises as a non-relativistic contraction \cite{Inonu:1953sp}
 of the  conformal algebra \cite{Jackiw:2000tz,Jackiw:2002he,Duval:2002cw,Lukierski:2005xy}. 
The contraction
amounts to taking a certain scaling limit of the generators. If these are Killing vectors of a metric, as in AdS/CFT,  then the contraction can be implemented in the metric. Indeed this is what happens in 
the Penrose limit: the contraction of the metric in this case  gives generically pp-wave space-times, while in the field theory side it corresponds to the scaling limit
discussed in \cite{Berenstein:2002jq}. In the context of non-relativistic AdS/CFT, an analogous scaling limit was performed in \cite{Maldacena:2008wh}. The String Theory side of this limit had been considered before\footnote{We thank the authors of \cite{Kruczenski:2008bs}
 for bringing their work to our attention.} in \cite{Kruczenski:2008bs}. 
However, the situation here is different: a scaling limit of the AdS metric does not give rise to a non-degenerate bulk metric realizing the symmetry. 
We instead perform a modified version of the contraction involving spin generators,
and we show that when applied to the Dirac equation
this produces non-trivial results in the limit. 
In particular, the centrally-extended algebra, 
and a free field wave equation, replacing the \sch\ equation, can be derived in this way. 
The interpretation of the exotic central extension as a non-relativistic limit of spin was first 
pointed out in \cite{Jackiw:2000tz,Jackiw:2002he,Duval:2002cw}.

We will study some generic features of conformal field theories 
with exotic  Galilean symmetry. As examples, 
we will present a free field and a Chern-Simons-matter-type action, invariant under the symmetry group, at least 
classically.  Following the strategy of \cite{Duval:1984cj,Duval:1990hj,Duval:2008jg} 
and of \cite{son,mcgreevy}, 
we will present a metric which realizes 
the exotic conformal Galilei group as its isometry group. Disappointingly, however, this metric has signature 
$(3,4)$, and it turns out to be a  ``wrong-signature'' versions of AdS$_7$, or its
deformation preserving the \sch\ group.  Therefore its
 use for AdS/CFT applications appears problematic. Despite this severe issue, we point out that 
a blind application of the AdS/CFT prescriptions yields a consistent result for the two-point functions.

The rest of the paper is organized as follows. In Sec.~\ref{sectwo} we briefly
 review the \sch\ algebra and other conformal Galilei algebras. 
We then discuss the central extensions of these algebras, and of the $l=1$ case in more detail. 
Sec.~\ref{contraction} discusses the origin of the algebra as a non-relativistic contraction. 
In Sec.~\ref{secthree} we study formal properties of field theories with exotic conformal Galilei algebras.
In Sec.~\ref{free-theory} we discuss explicit field theories examples. In Sec.~\ref{metricsec} 
we present  an invariant metric and comment on a holographic-type computation of two-point functions, based on this background metric. In Sec.~\ref{conclusions} we conclude. 
In Appendix \ref{lagappe} we discuss a ``particle'' Lagrangian  with
exotic conformal Galilei symmetry and its quantization.

\bigskip

{\bf Note}: while writing the present article, a paper discussing conformal Galilei algebras
appeared \cite{Bagchi:2009my}. Just before submitting, we became aware of 
two other related papers \cite{Alishahiha:2009np}, \cite{Bagchi:2009ca}. However, our main 
results have not been discussed in these references.

\section{Conformal Galilei algebras}
\label{sectwo}

\subsection{The \sch\ algebra}

We will begin by recalling the \sch\ group and its geometric realization. As we will see, the conformal Galilei groups 
that we will discuss in this paper are a natural generalization of the \sch\ one. The fundamental symmetry of 
a non-relativistic system is the Galilei group. A point in $d$-dimensional Euclidean space $\vec{x}$, 
at a given time $t$, transforms under the Galilei group as
\bea
 t \to t + e, \qquad \vec{x} \to R\vec{x} + \vec{b}t + \vec{c} 
\eea
where $R\in SO(d)$, $\vec{b},\vec{c}\in \R^d, e \in \R$.
Let us denote by $X$ the pair $(t,\vec x)$,
and denote by $P(\vec c)$ and $K(\vec b)$ the geometric actions of finite translation by $\vec c$
and finite boost by $\vec b$, respectively.

Quantum mechanically,
wave functions must form a projective unitary representation of the group, 
which is equivalent to a
unitary representation of the centrally extended Bargmann group \cite{Bargmann:1954gh}. 
Consider the free \sch\ equation with mass $m$,
\bea
\left(i \de_t  +\frac{1}{2m}\de_i \de_i \right) \psi = 0~.
\eea 
One can trivially promote the geometric action of the translation $P(\vec c)$
to an action $U_{P(\vec c)}$ on the wavefunction by defining \begin{equation}
U_{P(\vec c)}\psi(X)= \psi(P(-\vec c)X),
\end{equation} however the action $\psi(X)\to \psi(K(-\vec b)X)$ does not map
a solution to another solution. Defining the unitary operator
$U_{K(\vec b)}$ by further multiplying by a  phase which depends both on space and time,
\begin{equation}
U_{K(\vec b)}\psi(X)= \exp \left[-im\vec{b}\cdot \vec{x} +\frac{i}{2}mt\vec{b}^2\right]\psi(K(-\vec b)X)~,
\end{equation}
one finds that the plane wave solution  
$\psi_{E,\vec p} = \exp(iEt - i\vec{p}\cdot \vec{x})$ is mapped to another plane wave,
\begin{equation}
\psi_{E,\vec p} \to U_{K(\vec b) }\psi_{E,\vec p}=\psi_{E',\vec p{}\,'}
\end{equation} where \begin{equation}
E'=\frac{(\vec p + m\vec b)^2}{2m} , \qquad \vec p\,'=\vec p + m\vec b. 
\end{equation}
The operators  $U_{P(\vec c)}$ and $U_{K(\vec b)}$ no longer commute,
instead we have 
\begin{equation}
U_{P(\vec c)}U_{K(\vec b)}  = \exp(i m \vec b\cdot\vec c)
U_{K(\vec b)} U_{P(\vec c)},
\end{equation} i.e. they  realize  a central extension.
 One can add an extra coordinate $\xi$ and consider the wave function $\psi = \exp(im\xi + iEt - i\vec{p}\cdot \vec{x})$. This
is \emph{invariant}, provided $\xi$ transforms as 
\bea
\xi \to \xi + \vec{b} \cdot \vec{x} +\frac{1}{2}t\vec{b}^2~.
\eea
In turn, the \sch\ equation may be interpreted as the Klein-Gordon equation in the Bargmann space $\R^{d+1,1}$, 
with metric
\bea
\diff s^2 = 2\diff \xi \diff t + (\diff \vec{x})^2 ~.
\label{bargmetric}
\eea

In fact, the symmetry group of the \sch\ equation is not just the Bargmann group, but it comprises also 
dilatations, $D: (t,\vec{x}) \to (\lambda^2 t, \lambda \vec{x})$,
and special conformal transformations,
$C: (t, \vec{x}) \to (1+ft)^{-1}(t, \vec{x})$. 
The complete action of the 
\sch\ group on the extended space-time is
\bea
&&\vec{x} \to\;  \frac{R \vec{x} + \vec{b} t + \vec{c}}{ft+g}~, \qquad  t \; \to \; \frac{dt+e}{ft+g}, \label{l=1/2vec} \nn\\[2mm]
&&\xi  \to\;  \xi+ \frac{f}{2} \frac{(R \vec{x} + \vec{b} t + \vec{c})^2} {ft+g} - \vec{b}\cdot  R \vec{x} - \frac{t}{2} \vec{b}^2 + h
\label{bargs}
\eea
where $dg-ef =1$.  The corresponding algebra\footnote{The normalization of the $D$ here is $1/2$ of the usual normalization in the \sch\ algebra.} of generators is given by
\be
\begin{array}{lll}
~[D,H] = H~, & \qquad [D,C] =-C~, &\qquad [C,H]=2D~, \\
~ [H,P_i] = 0~, &\qquad  [D, P_i] = \frac{1}{2}P_i~, &\qquad [C,P_i]= K_i~ ,\\
~ [H,K_i] = -  P_i~, &\qquad  [D, K_i] = -\frac{1}{2}K_i~, &\qquad [C,K_i]= 0
\end{array} \label{sch-commutators}
\ee 
plus the central extension 
\begin{equation}
[P_i,K_j]=\delta_{ij} M~.
\label{centralm}
\end{equation} 
Momenta $P_i$ and boosts $K_i$  transform as vectors under spatial rotations $J_{ij}$, 
and  as a doublet  of the $SL(2;\R)$ sub-algebra generated by $H,D,C$. 
Based on the above remarks, 
it can be shown that the \sch\ algebra is a sub-algebra of the relativistic conformal algebra 
in one higher space-time dimension \cite{BurdetPerrinSorba}. This was indeed a key observation for the holographic  realization of the \sch\
symmetry \cite{son,mcgreevy}.

\subsection{``Spin-$l$'' conformal Galilei algebras}

We now  present a class of conformal extensions of the Galilei algebra labeled by a half-integer $l$, discussed in  \cite{negro1,negro2}. 
The smallest instance $l=1/2$ corresponds to the (non centrally-extended) \sch\ algebra. The next case, $l=1$,
is an extension of the Galilei group comprising constant \emph{accelerations}, and we will study it in more detail.
Below, we follow \cite{negro1,negro2} and present the algebras without central extensions. Those with central extensions
will be discussed in the next subsection.

The action on space-time may be written as
\bea
\vec{x}  \to  \frac{R \vec{x} + t^{2l}\vec{c_{2l}} +\dots + t^2 \vec{c_2} + t \vec{c_1}  + \vec{c_0}}{(ft+g)^{2l}}~, \qquad \quad
t  \to  \frac{dt+e}{ft+g}
\eea
where $R\in SO(d)$, $\vec{c}_{n}\in \R^{d}$ and $dg-ef=1$, generalizing the $l=1/2$ 
transformations \eqref{l=1/2vec}. 
From this we see that the infinitesimal action is generated by the following vector fields
\begin{align}
&& H &=  \de_t ,& D &= - t \de_t -l x_i \de_i,&  C &= t^2 \de_t +2 l t x_i \de_i, &&\nn\\
 && J_{ij} &= -(x_i \de_j - x_j \de_i),     &P_i^n &= (- t)^n \de_i ,  & &&&
\label{lcgal:oper}
\end{align} where $i=1,\ldots,d$ and $n=0,\ldots,2l$.
Notice that the dilatations act as 
\bea
t \to \lambda t ~,\qquad \vec{x}\to \lambda^{l} \vec{x}
\eea
corresponding to a dynamical critical exponent $z=1/l\leq 2$.

The algebra satisfied by the generators $\{H,D,C,J_{ij},P_i^{n}\}$, 
($n=0,\dots,2l$; $i=1,\ldots,d$)
comprises the following non-zero commutation relations 
\begin{equation}
\begin{array}{r@{}lr@{}lr@{}l}
[D,H] &= H, & [D,C]&=-C, & [C,H]&=2D, \\[1mm] \relax
[H,P^{n}_i]&=-nP_i^{n-1},  & [D,P^{n}_i] &= (l-n) P^{n}_i,    & [C,P^{n}_i]&= (2l-n)P_i^{n+1}, \\[1mm] \relax
  [J_{ij},P_k^{n}] &  = \delta_{ik} P_j^{n} - \delta_{jk} P_i^{n}, &  
\left[J_{ij},J_{kl}\right]  & \multicolumn{3}{l}{=  \delta_{ik} J_{jl} + \delta_{jl} J_{ik} - \delta_{il} J_{jk} - \delta_{jk} J_{il}  }.
\end{array}
\label{spinl}
\end{equation}
We can then take this as an abstract definition of the algebra, independent of the specific realization  \eqref{lcgal:oper}. Notice that  thes generators are anti-Hermitean, and 
$H,D,C$ form an $SL(2;\R)$ sub-algebra, familiar from the \sch\ algebra. The generators $P_i^{n}$ transform in the spin-$l$ representation of this $SL(2;\R)$, and as vectors of the $SO(d)$ sub-algebra of spatial rotations generated by $J_{ij}$ in the usual way. 

Setting  $l=1/2$ and defining $P_i^{0}=P_i$, $P_i^{1}=K_i$, we recover the \sch\ algebra \eqref{sch-commutators}, albeit without central extension.
For $l=1$, there is one extra set of generators, transforming as a vector of $SO(d)$,
that we will denote $P_i^{2}=F_i$. It generates accelerations $\vec{x} \to \vec{x}+ t^2 \vec{a}$. For future reference, let us  write down the non-trivial  commutators:
\be
\begin{array}{lll}
~ [H,P_i] = 0~, &\qquad  [D, P_i] = P_i~, &\qquad [C,P_i]= 2K_i~, \\
~ [H,K_i] = -  P_i~, &\qquad  [D, K_i] = 0~, &\qquad [C,K_i]= F_i~,\\
~ [H,F_i] = - 2 K_i~, &\qquad  [D, F_i] = -F_i~, &\qquad [C,F_i]= 0.
\end{array} \label{cgal-commutators}
\ee 
Notice that the dynamical exponent in this case is $z=1$, the same as the relativistic one. One interesting feature
of this algebra\footnote{In fact also the centrally-extended algebra, to be discussed momentarily.} 
is that it can be obtained as a \emph{non-relativistic contraction} of the relativistic conformal algebra \cite{Lukierski:2005xy}. We will discuss this  aspect later in Sec.~\ref{contraction}.

\subsubsection*{Infinite dimensional extension}

It is known that the \sch\ algebra admits an infinite dimensional Virasoro-like extension \cite{Henkel:1993sg,Henkel:2003pu}, which  
should be relevant for holographic applications \cite{Alishahiha:2009nm}. 
We notice that such Virasoro-like extensions
exist for every $l$. This is easily showed in the representation (\ref{lcgal:oper}). First, we rename the $SL(2;\R)$
generators as $L^{-1}=-H$, $L^{0}=D$, and $L^{1}=-C$, and then we define the following operators 
\begin{align}
L^n&=- t^{n+1}\partial_t - l (n+1) t^n x_i\partial_i ,\nn\\
P_i^n &= -t^{n+l}\partial_i ,\nn \\
J_{ij}^n &= -t^n (x_i\partial_j-x_j\partial_i)
\end{align} 
where $n\in \Z$ for the $L^n$, and $n\in \Z+l$ for the $P_i^n$. The generators
$\{P_i^{-l}, \dots, P_i^{l}\}$ are those of the finite-dimensional conformal Galilei algebras (after a relabeling $n\to n-l$ and multiplying by $\pm1$). 
The commutators are  then written in the compact form 
\begin{align}
[L^m,L^n]&=(m-n)L^{m+n},\nn\\
[L^m,J^n_{ij}]&= -n J^n_{ij},\nn \\
[J^m_{ij},J^n_{kl}]&=  \delta_{ik}J^{m+n}_{jl}+\delta_{jl}J^{m+n}_{ik}-\delta_{il}J^{m+n}_{jk}
-\delta_{jk}J^{m+n}_{il},\nn \\
[L^m,P^n_i] &= (lm-n) P_i^{m+n},\nn \\
[J_{ij}^m,P^n_k] &= \delta_{ik}P_j^{m+n}-\delta_{jk}P_i^{m+n}.
\end{align} 
The $L^n$ generators form a Virasoro algebra,
$J^n_{ij}$ is an $so(d)$ current algebra,
and $P^n_i$ give  weight-($l+1$) Virasoro primaries
which are also primary under the current algebra;
they do not have any central extension at all at this stage.
One can add the usual central 
extension $\delta_{n+m,0}(n^3-n)c/12$ to the $[L^m,L^n]$ commutators,
which might arise quantum mechanically, or classically as in \cite{Brown:1986nw}
if we have a holographic dual. 

\subsection{Central extensions}
\label{centsec}

As we reviewed, the Galilei group admits the central extension \eqref{centralm}, 
which was originally found by Bargmann \cite{Bargmann:1954gh}.
This central extension  plays a crucial role in realizing geometrically \cite{Duval:1984cj} and holographically \cite{son,mcgreevy}
the \sch\ symmetry. 
Motivated by this, we will now show that there exists a
central extension of the spin-$l$ conformal Galilei groups. 
The  algebras in the previous sub-section are of the form  
$\frakg\oplus V $,
where $\frakg$ is a Lie algebra and $V$ its representation.
In our case $\frakg$ is $sl(2;\R)\oplus so(d)$ 
and $V$ is the tensor product of the spin-$l$ representation of $SL(2;\R)$ and
the vector representation of $SO(d)$.
The commutator of $g\in \frakg$ and $v\in V$ is defined by \begin{equation}
[g,v] = gv,
\end{equation} where the right hand side stands for the action of $g$ on $v$.
For $v,w\in V$ we have \begin{equation}
[v,w]=0.\label{without-central}
\end{equation} Since $\frakg$ is semisimple,
the  central extension can only enter  in \eqref{without-central}  as\begin{equation}
[v,w]=K(v,w)
\end{equation} where $K(v,w)$ is an antisymmetric pairing. 
The Jacobi identity is then satisfied if and only if $K(v,w)$ is invariant under $\frakg$.
Now, the invariant tensor $I^{mn}$ of the spin-$l$ representation of $SL(2;\R)$ is 
symmetric or antisymmetric depending on the parity of $2l$,
whereas $SO(d)$ has an antisymmetric invariant tensor $\epsilon_{ij}$ for $d=2$
and a symmetric tensor $\delta_{ij}$ for any $d$.

From these considerations, we conclude that there are two types of central extension of
conformal Galilei groups.
One exists for any $d$ and half-integral $l$: 
\begin{equation}
[P^m_i,P^n_j]= I^{mn} \delta_{ij} M~,
\end{equation} 
and another which exists only for $d=2$ and integral $l$: 
\begin{equation}
[P^m_i,P^n_j]= I^{mn} \epsilon_{ij} \Theta~.
\end{equation}
Let us discuss the two cases which we are most interested in. 
For arbitrary $d$ with $l=1/2$, one has the massive central extension \eqref{centralm} of the
algebra  in \eqref{sch-commutators}.
This is the familiar \sch\ symmetry with the Bargmann extension.
In order to realize the centrally extended algebra by the action of vector fields,
one needs to add one extra direction $\xi$ which is 
conjugate to $M$. Equivalently, one can Fourier-transform with respect to the mass $M$.
The generators take the form
\bea
\begin{array}{lll}
H = \partial_t, &   \quad D= - \frac12x_i \partial_i - t \partial_t, &\quad
 C=  t x_i \partial_i  + t^2 \partial_t   - \frac12x_ix_i \partial_\xi,\\[2mm]
P_i =\partial_i, & \quad K_i=-t \partial_i  - x_i \de_\xi, &\quad  J_{ij} = -  (x_i \de_j - x_j \de_i),
\end{array}
\eea
and $M= \partial_\xi$.

Next consider $d=2$ with $l=1$. 
The centrally extended algebra has commutators
\bea
[K_i, K_j ] = \Theta\epsilon_{ij}~, \qquad [P_i, F_j ] = -2\Theta\epsilon_{ij}
\eea 
in addition to those in \eqref{cgal-commutators}. We denote the eigenvalue of 
$\Theta$ by $i\theta$.
Following \cite{Duval:2000xr,Lukierski:2005xy},
we will refer to the resulting symmetry algebra as the {\em exotic conformal Galilei algebra} 
which will be the focus of the rest of our paper.
The fact that the two-dimensional Galilei group admits a second central extension $\Theta$,
in addition to the mass parameter $M$ has long been known \cite{LevyLeblond,Grigore:1993fz,Bose:1994sj}, and
various systems have been discussed in the literature \cite{Lukierski:1996br,Stichel:2003kh,delOlmo:2005md,Lukierski:2007nh}. 
We see that the existence of the extra conformal generators only allows $\Theta$, not $M$,
as noted in \cite{Lukierski:2005xy}.

We can represent this algebra by modifying the operators in
\eqref{lcgal:oper} with extra terms which produce the central extension. 
Concretely, we have
\bea
\begin{array}{lll}
H= \partial_t, & \qquad  D= - x_i \partial_i - t \partial_t, &\qquad
C= 2 t x_i \partial_i  + t^2 \partial_t   - 2x_i\chi_i,\\[2mm]
P_i= \partial_i, & \qquad
K_i=-t \partial_i  + \chi_i & \qquad
F_i= t^2 \partial_i - 2t \chi_i + 2 x_j \epsilon_{ij}\Theta~,
\end{array}
\label{gengen}
\eea
where we require that the operators $\chi_{1,2}$ and $\Theta$ satisfy
the \emph{Heisenberg} commutation relations
\begin{equation}
[\chi_i,\chi_j]=i\Theta \epsilon_{ij}, \qquad\quad  [\chi_i,\Theta]= 0~,
\label{heisalg}
\end{equation} 
and that they commute with the operators defined in \eqref{lcgal:oper}.
The action of the modified generators can be seen as the sum of the geometric action \eqref{lcgal:oper} of the group
without the central extension,  modified by the part proportional to $\chi_i$ and $\Theta$.

Let us now discuss the rotation generator.
First of all, notice that in order for the generators $P_i,K_i,F_i$ to transform as vectors under rotations,  
$\chi_i$ should transform as vectors. Then the rotation generator comprises two pieces:
\bea
J_{ij} =  -(x_i \de_j - x_j \de_i )+ S_{ij}
\eea
where $S_{ij}$ rotates the $\chi_i$.
Inspired by the observation in \cite{Lukierski:2007nh},  we define the rotation generator as 
$2i\theta J \equiv F_i P_i - K_i K_i$, so that 
\begin{equation}
J = -\epsilon_{ij} x_i \frac{\partial}{\partial x_j}  - \frac1{2i\theta} \chi_i\chi_i~.
\label{fancy-rotation}
\end{equation}
With this definition we have 
\begin{equation}
[J,x_i]=\epsilon_{ij}x_j, \qquad [J,\chi_i]=\epsilon_{ij}\chi_j
\end{equation} 
thus this $J$ correctly transforms all of $P_i$, $K_i$ and $F_i$ as vectors.
It is clear that we can think of $\chi_i$ as an intrinsic part of the Galilei boost,
just as the rotation generator acting on a field with non-trivial spin is not simply
$x_{[i} \partial_{j]}$ but $x_{[i}\partial_{j]} + \Sigma_{ij}$.
We will pursue this analogy further in Sec.~\ref{contraction}: this form naturally arises
by taking the contraction of Lorentz generators acting on spinor fields.

We can obtain a realization of the algebra in terms of \emph{vector fields}, by representing the Heisenberg 
algebra in an auxiliary three-dimensional space with coordinates 
$v_{1}, v_2$, and $\xi$. In particular, we can take
\begin{equation}
\Theta=\partial_\xi, \qquad \quad \chi_i= \frac{\partial}{\partial v_i} - \frac12 \epsilon_{ij}v_j\partial_\xi~.
\label{vrep}
\end{equation} 
This is a generalization of the Bargmann space to the present case. 
We will study natural metrics on this space later. 
The algebra of vector fields is obtained by just plugging 
\eqref{vrep} into \eqref{gengen}. For the rotation generator, we have
\bea
J  = -\epsilon_{ij}(x_i \frac{\de}{\de x_j} + \frac{1}{2}v_i \frac{\de}{\de v_j}) - \frac{1}{2i\theta}\frac{\de}{\de v_i}\frac{\de}{\de v_i} - \frac{i\theta}{8}v_iv_i~.
\label{jvv}
\eea
Notice that this does not rotate independently $\de /\de v_i$ and $v_i$.
Actually, in this representation the naive definition of the internal rotation generator 
\bea
S_{ij} = - (v_i \frac{\de}{\de v_j} - v_j \frac{\de}{\de v_i})~,
\label{vrot}
\eea
gives the  correct commutation relations, thus both choices are possible in this case. We will see however that 
the more general definition \eqref{fancy-rotation} arises naturally from the contraction procedure.  Notice
that the internal part of the rotation operator \eqref{fancy-rotation} is a \emph{harmonic oscillator} Hamiltonian.

The induced infinitesimal 
action on the extended six-dimensional space 
with coordinates $(t,x_i,\xi,v_i)$ is easily obtained from the vector fields, and reads
\begin{align}
H&: t \to t + e,\label{htra}\\
D&: t \to t -\lambda t, \quad  x_i \to x_i - \lambda x_i,\\
C&:  t \to t+ a t^2,  \quad  x_i \to x_i + 2at x_i, \nn\\
& \;\;\,  v_i \to v_i - 2a x_i , \quad \xi \to \xi + a \epsilon_{ij} x_i v_j,\label{l1conf}\\
P_i&: x_i \to x_i + c_i,\\
K_i&: x_i \to x_i - b_i t, \quad \;  v_i \to v_i + b_i, \qquad \xi\to \xi -\frac{1}{2} \epsilon_{ij}b_i v_j,\label{l1boost}\\
F_i&: x_i \to x_i + a_i t^2, \quad v_i \to v_i -2a_i t, \quad \xi \to \xi + \epsilon_{ij}a_i (2x_j+tv_j),\label{l1acc}\\
\Theta& :\xi \to \xi + h~.\label{ttra}
\end{align}
Notice the transformations of the $v_i$ may be  interpreted as (minus) 
the transformations of \emph{velocities} $\dot{x}_i$. 
A general transformation of $\xi$,
\bea
\xi \to \xi + a \epsilon_{ij} x_i v_j + \epsilon_{ij}a_i (2x_j+tv_j)
 -\frac{1}{2} \epsilon_{ij}b_i v_j + h \label{xi-shift}~,
\eea 
at first sight does not have an obvious interpretation. 
However it  has a natural interpretation in terms quasi-invariance of a free particle mechanical system that we discuss in Appendix \ref{lagappe}.

\section{Exotic conformal Galilei algebra from contraction}
\label{contraction}

One interesting aspect of the $l=1$ conformal Galilei algebra is that  
it can be obtained as the contraction of the conformal algebra $SO(d+1,2)$, 
just as the Galilei algebra
arises as the contraction of the Poincar\'e algebra \cite{Inonu:1953sp}.
 This raises the hope that applying  the contraction to 
 the AdS$_{d+2}$ metric, we could obtain a non-trivial metric with desired
isometries. However, it is easy to see that such hope cannot be realized, as there is no $d+2$ dimensional metric which is non-degenerate
with such isometries. The reasoning for this is analogous to that showing that there is no $d+1$ dimensional metric, with ordinary Galilei symmetries. 
Indeed, performing the limit in the metric gives a degenerate result.

\subsection{Improved contraction}

In this subsection we will discuss a modified contraction that leads to the centrally extended conformal Galilei algebra in $d=2$ dimensions. The role of spin in the contraction was emphasized in 
\cite{Jackiw:2000tz,Jackiw:2002he,Duval:2002cw}, 
however our treatment here is slightly different, in particular we show that the Heisenberg algebra emerges
in the limit\footnote{We have recently learned that this contraction was considered previously in \cite{Horvathy:2004fw}, albeit the discussion there concerns a non-conformal algebra, with two central extensions $\Theta$ \emph{and} $M$.}. 
We start from an explicit representation of the relativistic conformal algebra, where  the generators of $SO(3,2)$ act on operators with spin, and perform a simultaneous contraction of the space-time and spin parts. Namely, we take 
\begin{align}
&&\tilde M_{\mu\nu} & =   x_\mu \de_\nu - x_\nu \de_\mu+ \Sigma_{\mu\nu}, &   \tilde  P_\mu  &=  \de_\mu, &&\nn\\
&&\tilde K_{\mu} & =  x^\nu x_\nu \de_\mu - 2 x_\mu x^\nu \de_\nu - 2x^\nu \Sigma_{\mu\nu},  
& \tilde D &=  x^\nu\de_\nu~, &&
\end{align}
where the  $\Sigma_{\mu\nu}$ can be thought of as a finite dimensional representation of $SO(1,2)$.
We then set $x_0=ct$ and  after rescaling the spin generators as
\bea
\Sigma_{i0} \to c \Sigma_{i0}, \qquad  \Sigma_{ij} \to c^2 \Sigma_{ij}~,
\eea
we define  the following generators 
\begin{align}
&&&& P_i &= \tilde P_i,   & K_i & = \frac{\tilde M_{i0}}{c}, &  F_i& = \frac{\tilde K_i}{c^2},    &&&& \nonumber  \\
&&&& H & = \tilde P_{0}, &  D & =\tilde D,   
   & C &= \frac{\tilde K_{0}}{c}, &&&&
\end{align}
as well as
\bea
 J= - \tilde M_{12} + c^2\Sigma_{12}~.
 \label{rotm}
\eea 
Taking  the $c\to \infty$ limit  we find the following operators
\bea\everymath{\displaystyle}
\begin{array}{lll}
H= \partial_t,\quad & D= - t \partial_t - x_i \frac\partial{\partial x_i} ,\quad& C=  t^2 \partial_t +   2 t x_i \frac\partial{\partial x_i} + 2 x_i \Sigma_{i0},\\[2mm]
P_i=\frac\partial {\partial x_i},\quad &
K_i=-t \frac\partial{\partial x_i} - \Sigma_{i0},\quad &
F_i= t^2 \frac\partial{\partial x_i} +2t \Sigma_{i0} +2 x_j \Sigma_{ij},
\label{sigmabasis}
\end{array}
\eea
and $\Theta_{ij}= \Sigma_{ij}$.
We will address rotations momentarily.  
The scaling limit 
of the commutation relations among the $\Sigma_{\mu\nu}$ gives rise to
\bea
 [\Sigma_{0i},\Sigma_{0j}] = \Sigma_{ij}, \qquad [\Sigma_{ij},\Sigma_{0k}] = 0 ~.
\eea 
Defining  $\chi_i = \Sigma_{0i}$, $\Sigma_{ij}= \epsilon_{ij}\Theta$, we see that these three operators obey precisely the Heisenberg commutation relations  (\ref{heisalg}).
Indeed it is well known that the Heisenberg group is a contraction  of the $SU(2)$ group\footnote{We could also consider a different contraction, giving  the Euclidean group, 
but this turns out to be not consistent.}. 
One can pick  a representation of the Heisenberg algebra,
where the $\chi_i$ take a concrete form. For example, one could consider a finite dimensional 
representation in terms of $3\times 3$ matrices, but this is not unitary. 
An unitary representation is of course 
given by $\chi_1 = s, \chi_2=i\theta \de_s$.
As we discussed in Sec. \ref{centsec}, a unitary representation in terms of vector fields is 
obtained introducing three auxiliary coordinates  $\xi,v_1,v_2$ and defining (\ref{vrep}).

Let us now address the rotations.
Firstly, notice that with the definition (\ref{rotm})
one result of the $c\to \infty$ limit of the commutation relations of the 
Lorentz rotations $[\tilde M_{\mu\nu},\tilde M_{\mu\nu}]$  is that 
\bea
~[J_{ij},J_{k0}+\Sigma_{k0}] & = & \delta_{ik} (J_{j0}+\Sigma_{j0}) - \delta_{jk} (J_{i0}+\Sigma_{i0}) ~. 
\label{problem}
\eea
While one can  check that all other components give a limit which is consistent with the contraction
of the $SO(1,2)$ algebra to the Heisenberg one, we see that 
 (\ref{problem}) is problematic. In particular, 
the spatial rotations $J_{ij}$ do not act on $\Sigma_{0k}$. Thus, the naive contraction procedure 
is not consistent. 

To obtain a consistent contraction we apply the following trick. 
We  consider a trivial central extension of the relativistic algebra, obtained adding a constant to the 
relativistic spatial rotations:
\bea
  - \epsilon_{ij}x_i \de_j  \to   - \epsilon_{ij}x_i \de_j + \alpha (2\Sigma_{0i}\Sigma_{0i}- \Sigma_{ij}\Sigma_{ij}) +\beta~.
\eea
Then the contraction limit 
is consistent, provided we take $\alpha = i/(4\theta c^2)$ and $\beta=- i\theta c^2/4$. 
The rotation generator in the limit is indeed \eqref{fancy-rotation}. In particular, the divergent term 
$c^2\theta$ that needs to be subtracted to get a finite answer in the limit, is analogous to the divergent rest mass $c^2M$. 
We see that  the parameter $\theta$  is a remnant  of spin in the non-relativistic limit.

\subsection{Contraction of the Dirac equation}

The relevance of  spin in the non-relativistic limit discussed in the previous subsection suggests that one
should be able to take an appropriate limit of a relativistic equation for fields with spin, for example the Dirac equation.  Notice in particular that the non-relativistic limit of the Klein-Gordon equation 
(in flat space) is degenerate.

Let us then  start from the massless Dirac equation in $2+1$ dimensions:
\bea
\Gamma^\mu \de_\mu \Psi = (-\Gamma^0 \de_t + \Gamma^1 \de_1 +\Gamma^2 \de_2 ) \Psi = 0 
\eea
where $\Gamma^a$ are gamma-matrices of $SO(1,2)$.
To take the limit, it is convenient to multiply by $\Gamma^{012}$:
\bea
(\Gamma^{12}\de_t - \Gamma^{02} \de_1 +\Gamma^{01} \de_2 ) \Psi = 0~. 
\eea
Then rescaling as in the previous subsection, we see that the equation just acquires an overall factor of $c$.
However, we should replace the original  gamma matrices with the operators obeying the Heisenberg commutation relations.
Then we immediately obtain 
\bea
(\Theta \de_t + \epsilon_{ij} \chi_{i} \de_j ) \Psi = 0 ~,
\eea
which in operator notation takes the form
\bea
(\Theta H +\epsilon_{ij} K_{i} P_j ) \Psi = 0 ~.
\label{MTeqn}
\eea
Non-relativistic limits of the Dirac equation have been studied in the literature. In particular, 
two well-studied equations are the Pauli equation and the  L\'evy-Leblond equation \cite{LLeq}. 
However, the equation \eqref{MTeqn} is not equivalent to these\footnote{As far as we are aware,
an equation closely related to \eqref{MTeqn} was presented in \cite{Horvathy:2004fw}. However, the discussion 
there assumes a non-zero mass $m$.}. In particular, notice that this is an infinite-component equation.

We will see in Sec.~\ref{secthree} that the same equation arises by considering a field
which saturates the unitarity bound.
This is analogous to the \sch\ case, where fields saturating the unitarity bound obey the free \sch\ equation, which in turn arises from the DLCQ of the Klein-Gordon equation.  We regard equation 
\eqref{MTeqn} as the counterpart of the free \sch\ equation in the present case, and we will discuss it 
further in Sec.~\ref{free-theory}.

\section{Properties of CFTs with exotic conformal Galilei symmetry}
\label{secthree}

\subsection{Action on operators}

Local operators ${\cal O}(t,\vec{x})$ depending on position in time and space are defined as
\bea
{\cal O}(t,\vec{x}) = \ex^{-Ht - P_ix_i} {\cal O}(0)\ex^{Ht +P_ix_i}~.
\eea
Such operators may be labeled by their scaling dimension and $\Theta$-quantum 
number
\begin{align}
~[{\cal O},D] & =   \Delta_{\cal O} {\cal O},&
~[{\cal O},\Theta] & =  i \theta_{\cal O} {\cal O},
\end{align}
and in general form a representation of the exotic conformal Galilei algebra. Notice that $[{\cal O},P_i]$ and  $[{\cal O},H]$ have scaling dimension $\Delta_{\cal O}+1$, $[{\cal O},K_i]$ has unchanged scaling dimension $\Delta_{\cal O}$, while
$[{\cal O},F_i]$ and $[{\cal O},C]$ have  scaling dimensions $\Delta_{\cal O}-1$. Thus it is natural to define
as \emph{primary} operators those satisfying 
\bea
~[{\cal O},F_i]  = [{\cal O},C]  = 0~.
\eea
From a primary operator it is possible to construct a tower of operators by acting with $H$, $P_i$, and $K_i$,
forming an irreducible representation of the algebra. 

When the operator $\cO$ is inserted at a generic point, one finds the following
action of the exotic conformal Galilei symmetry: 
\be
\begin{array}{ll}
 [\cO,H] =\partial_t \cO, & [\cO,P_i] = \frac{\partial}{\partial x_i}\cO, \\[2mm]
 [\cO,D] =\left(-t\frac{\partial}{\partial t} - x_i\frac{\partial}{\partial x_i} -\Delta\right) \cO, &
[\cO,K_i] = \left(-t \frac{\partial}{\partial x_i} + \chi_i\right)\cO, \\[2mm]
[\cO,C] =\left(t^2\frac{\partial}{\partial t}   +2t x_i \frac{\partial}{\partial x_i}
-2 x_i \chi_i+2t\Delta \right)\cO, & [\cO,F_i] = \left(t^2 \frac{\partial}{\partial x_i} -2t \chi_i + 2i\theta x_j \epsilon_{ij}\right) \cO,\\[2mm]
[\cO,J] = \left( -\epsilon_{ij} x_i \frac{\partial}{\partial x_j}  - \frac1{2i\theta} \chi_i\chi_i\right) \cO~. & \label{action-on-fields}
\end{array} 
\ee

Note that the operator $\hat K_i=K_i + t P_i$ commutes with $H$ and 
$P_i$ once we identify $H$ with $\partial_t$ as above. Then 
we can introduce extra directions $v_{1,2}$ by defining
\begin{align}
{\cal O}(t,\vec x,\vec v) &= \ex^{-\hat K_i v_i} 
{\cal O}(t,\vec x)\ex^{  \hat K_i v_i}~.
\label{locoper}
\end{align} 
Let us see how the generators  act in this representation.
Using the Zassenhaus formula, we have
\begin{align}
\cO(t,\vec x,\vec v) &=
\ex^{i\theta v_1v_2/2} \ex^{-\hat K_1v_1}\ex^{-\hat K_2v_2}
\cO(t,\vec x)
\ex^{\hat K_2v_2}\ex^{\hat K_1v_1}.
\end{align} 
Then, taking the derivative with respect to $v_i$, we get
\begin{equation}
[\cO(t,\vec x,\vec v),\hat K_i]=
\left(\frac{\partial}{\partial v_i} - \frac{i\theta}2  \epsilon_{ij}v_j\right) \cO(t,\vec x,\vec v )~.
\end{equation} 
This shows that on local operators \eqref{locoper}
the action of $\chi_i$  in \eqref{action-on-fields} is given by 
\bea
\chi_i = \frac{\de}{\de v_i} - \frac{i\theta}{2} \epsilon_{ij} v_j~.
\eea

\subsection{State-operator correspondence}

It is well-known that  in relativistic conformal field theories there exists 
a one-to-one correspondence between operators and states. One way to establish this is to perform radial quantization.
In \cite{Nishida:2007pj} an analogous correspondence was worked out for non-relativistic CFTs with \sch\ symmetry. 
As we show below, this construction generalizes straightforwardly to higher $l$, because it is based
on the common $SL(2;\R)$ sub-algebras. For simplicity we will concentrate on the $l=1$ case.

We define a state $\ket{\Psi_{\cal O}}$ associated to the operator ${\cal O}$ via the formula
\bea
\ket{\Psi_{\cal O}} = \ex^{\frac{\pi i }{4}(C-H)}{\cal O}\ket{0} 
\label{defstate}
\eea
where $\ket{0}$ is the vacuum, defined as a state which satisfies
\bea
G \ket{0} = 0 
\eea 
for all generators $G$.
Here we think of ${\cal O}$ as a creation operator. 
If ${\cal O}$ is a  primary, this formula reduces to 
\bea
\ket{\Psi_{\cal O}} = \ex^{-\frac{\pi i }{4}H}{\cal O}\ket{0} 
\eea which was used in \cite{Nishida:2007pj}.
However (\ref{defstate}) holds more generally. Using the relation
\bea
\tilde{D}\equiv\ex^{\frac{\pi i }{4}(C-H)} D \ex^{-\frac{\pi i }{4}(C-H)} = \frac{i}{2} (H+C)
\label{Dtilde}
\eea
it is easy to check that $\ket{\Psi_{\cal O}}$ has definite scaling dimension, namely it obeys
\bea
\tilde D \ket{\Psi_{\cal O}} = i \Delta_{\cal O}\ket{\Psi_{\cal O}}~.
\eea
Since the angular momentum $J$ commutes with dilatations $D$, we can label  
highest weight states as $\ket{\theta,\Delta,j}$, where $j$ is the angular momentum quantum number. 
In the \sch\ case such representations have been discussed in detail in \cite{Nakayama:2008qm} and are similarly characterized by $\ket{m,\Delta,j}$.

\subsection{Unitarity bound}

It is now easy to determine a lower bound on the scaling dimension of highest weight states imposed by unitarity.
For the \sch\ case this analysis has been done in \cite{Nakayama:2008qm,Lee:2009mm}. We do the analysis explicitly for the $l=1$ case, but the method we employ is applicable more  generally. In particular, we use the transformation (\ref{Dtilde}) to construct a non-unitary representation 
of the algebra, in terms of new generators 
\bea
\tilde G \,=\, \ex^{\frac{\pi i }{4}(C-H)} \, G\, \ex^{-\frac{\pi i }{4}(C-H)} ~.
\eea
This is an automorphism of the algebra which manifestly preserves the commutation relations,
but it changes their Hermiticity properties. Let us record the 
 transformed generators
\begin{align}
  \tilde H &=\frac12(H-C+2iD), & \tilde{D} &= \frac{i}{2} (H+C),    &  \tilde C &=\frac12(-H+C+2iD),\nonumber\\
 \tilde P_i &= \frac{1}{2}(P_i - F_i + 2iK_i), & \tilde K_i  &= \frac{i}{2}(P_i + F_i),  &\tilde F_i &= \frac{1}{2}(-P_i + F_i + 2iK_i)~.
\end{align} 
Notice that $\tilde K_i^\dagger = \tilde K_i$,  $\tilde F_i^\dagger = \tilde P_i$ and $\tilde \Theta^\dagger = -\Tilde \Theta$. 
Now we consider the state 
\bea
\ket\psi = (c\tilde\Theta \tilde H + \epsilon_{ij}   \tilde K_i \tilde P_j) \ket\phi
\eea
for an arbitrary real constant $c$. We also take $\ket\phi$ to be a primary, namely
\bea
\tilde F_i \ket\phi = \tilde C \ket\phi = 0 ~.
\eea
A short computation than gives the norm
\bea
\bra\psi \psi \rangle =  \theta^2 \bra\phi \phi \rangle 2(c^2\Delta -2c+1)\geq 0 ~.
\eea
We find the scaling dimension must satisfy
\bea
\Delta \geq  \frac{2c-1}{c^2}~.
\eea
The right hand side is maximized at $c=1$, thus we get the bound 
\bea
\Delta \geq 1 ~.
\label{l=1bound}
\eea
States $\ket{\psi_\mathrm{free}}$ saturating this bound then obey the equation
\bea
(\tilde \Theta \tilde H + \epsilon_{ij}\tilde K_i \tilde P_j ) \ket{\psi_\mathrm{free}}~= ~0, 
\label{stateeq}
\eea
which is the analogue of a free field obeying the \sch\ equation in the $l=1/2$ case.  
This corresponds to a field $\psi_\text{free}$ satisfying \begin{equation}
(\Theta H + \epsilon_{ij} K_i P_j) \psi_\text{free}~=~0~.
\end{equation} 
Remarkably this is exactly the equation \eqref{MTeqn} obtained from the 
improved contraction of the Dirac equation.
We will study
the free field theory associated to this equation in Sec.~\ref{free-theory}.

\subsection{Two-point functions}
\label{2pt-func}

In relativistic CFTs as well as non-relativistic CFTs, 
correlation functions are constrained by Ward identities.  
The form of correlation  functions of non-relativistic CFTs 
with \sch\ invariance was analyzed in
\cite{Henkel:1993sg}. For example, two-point 
functions are determined uniquely to be of the form
\bea
\langle {\cal O}_1(t,\vec{x}), {\cal O}_2(0,\vec{0})\rangle  = C \delta_{\Delta_1,\Delta_2} \delta_{M_1,M_2}\theta(t) t^{-\Delta_1}  \exp \left (i\frac{M_1}{2}\frac{\vec{x}^2}{t}\right)
\label{sch2pt}
\eea
where $C$ is a constant. To have a non-zero two-point function, the scaling dimensions and masses of the two local operators must be equal: $M_1=M_2$ and $\Delta_1=\Delta_2$.

Here we will  repeat the analysis for the case at hand, namely the $l=1$ exotic 
conformal Galilei  symmetry. 
We follow closely the exposition of \cite{Henkel:1993sg}. 
We consider the two-point function of local operators \eqref{locoper} in the extended space-time
\bea
F = \langle {\cal O}_1(t_1,\vec{x}_1,\vec{v}_1){\cal O}_2(t_2,\vec{x}_2,\vec{v}_2) \rangle~.
\eea
We then consider in turn the constraints imposed by the symmetries, where the action of the operators on $F$
is given by the sum of the actions  in \eqref{action-on-fields}, specified in the representation \eqref{vrep}. 
It is convenient to  define the variables
\begin{align}
t&   = t_1-t_2, & \vec{x}& =\vec{x}_1- \vec{x}_2,  &\vec{v}& =\vec{v_1}-\vec{v_2}, \nn\\
t_+& = t_1+t_2, &\vec{x}_+& =\vec{x}_1+ \vec{x}_2, &\vec{v}_+&=\vec{v_1}+\vec{v_2}~.
\label{newvar}
\end{align}
First of all, time and space translations imply that the correlation functions must 
be functions of $t,\vec{x}$, but in principle also of both $\vec{v},\vec{v}_+$.
Dilatation invariance then implies 
\bea
\left(t \frac{\de}{\de t} + x_i\frac{\de}{\de x_i} +\Delta_+ \right) F(t,\vec{x},\vec{v},\vec{v}_+) = 0
\eea
where $\Delta_+=\Delta_1 + \Delta_2$.
Defining $\vec{u}=\vec{x}/t$, we find that 
\bea
F(t,\vec{x},\vec{v},\vec{v}_+) = t^{-\Delta_+} G(\vec{u},\vec{v},\vec{v}_+)~.
\eea
Invariance under boosts gives
\bea
\left(   \frac{\de}{\de u_i}  - 2 \frac{\de}{\de v_{+i}} + \frac{i}{4}\epsilon_{ij}(v_{+j}\theta_+ + v_{j}\theta_-) \right) G(\vec{u},\vec{v},\vec{v}_+) = 0
\label{prewardboost}
\eea
where $\theta_{\pm} = \theta_1 \pm \theta_2$.
Now, let us consider the action of accelerations $F_i$. 
Changing variables, and hitting the equation with $\de/\de x_{+i}$, we find
that $\theta_+ G(\vec u,\vec v,\vec v_+)=0$, namely the ``selection rule''
\bea 
\theta_2 \, =\, -\theta_1~.
\eea
This is analogous to $M_1=M_2$ in the Bargmann group \cite{Henkel:1993sg}. 
Substituting this  back in (\ref{prewardboost}), 
we then find the following Ward identity
\bea
\left(  \frac{\de}{\de u_i}  - 2 \frac{\de}{\de v_{+i}} + i\frac{\theta_1}{2}\epsilon_{ij}v_{j} \right) G(\vec{u},\vec{v},\vec{v}_+) = 0~.
\label{ward:one}
\eea
Plugging this into the equation coming from accelerations we determine a second Ward identity
\bea
\left(  \frac{\de}{\de v_i}  - i\theta_1\epsilon_{ij}(u_j +\frac{1}{4}v_{+j} )\right) G(\vec{u},\vec{v},\vec{v}_+) = 0~,
\label{ward:two}
\eea
which can be immediately integrated to give 
\bea
 G(\vec{u},\vec{v},\vec{v}_+) = g (\vec{u},\vec{v}_+) \ex^{i\theta_1\epsilon_{ij} v_i (u_j +\frac{1}{4}v_{+j} )}~.
\eea
Plugging this into (\ref{ward:one}) we get 
\bea
\left(\frac{\de}{\de u_i} - 2 \frac{\de}{\de v_{+i}} \right) g (\vec{u},\vec{v}_+) = 0~.
\eea
Therefore we have
\bea
 G(\vec{u},\vec{v},\vec{v}_+) = g (\vec{u} + \frac{\vec{v}_+}{2}) \ex^{i\theta_1\epsilon_{ij} v_i (u_j +\frac{1}{4}v_{+j} )}~. 
\label{2pt-G}
\eea
Next, requiring invariance of $F(t,\vec x,\vec v,\vec v_+)$ 
under the conformal transformation $C$ and using  (\ref{ward:one}) we obtain
\bea
\left(- 2 x_i \frac{\de}{\de v_i}  + i\frac{\theta_1}{2} \epsilon_{ij}x_i v_{+j} + t (\Delta_1 - \Delta_2)\right) G(\vec u,\vec v,\vec v_+) = 0 
\eea
and contracting (\ref{ward:two}) with $x_i$ we find that the only new constraint it imposes is
\bea
\Delta_1 = \Delta_2 ~.
\eea

It remains to check rotation invariance. 
The rotation generator acting on one operator is given in \eqref{fancy-rotation}.
When acting on two-point functions, we have
\bea
J  = J_1 + J_2 =  - \epsilon_{ij}\left(x_i \frac{\de}{\de x_j}  + \frac12 v_i \frac{\de}{\de v_j} +\frac12 v_{+i} \frac{\de}{\de v_{+j}}\right) +\frac{2i}{\theta} \frac{\de}{\de v_i}\frac{\de }{\de v_{+i}} - i \frac{\theta}{8} v_iv_{+i}~,
\eea which is \begin{equation}
J= \ex^{i\theta_1\epsilon_{ij} v_i (u_j +\frac{1}{4}v_{+j} )}
\left(
-\epsilon_{ij}u_i(\frac{\partial}{\partial u_j}-2\frac{\partial}{\partial v_{+j}})
-\epsilon_{ij}v_i\frac{\partial}{\partial v_j}
+\frac{2i}{\theta}\frac{\partial}{\partial v_i}\frac{\partial}{\partial v_{+i}}
\right)
\ex^{-i\theta_1\epsilon_{ij} v_i (u_j +\frac{1}{4}v_{+j} )}.
\end{equation}
Thus we find that \eqref{2pt-G} is invariant under this.

In conclusion, we have proved that the general two-point function is 
\bea
F = \delta_{\Delta_1,\Delta_2} \delta_{\theta_1,-\theta_2} t^{-2\Delta_1} g (\vec{u} + \frac{\vec{v}_+}{2}) \exp \left(i\theta_1 \vec{v} \times (\vec{u} +\frac{1}{4}\vec{v}_+ )\right)~
 \label{2pt-func-final}
\eea
where we used a shorthand notation $\vec a\times \vec b\equiv \epsilon_{ij} a_i b_j$.
 Notice the 
$t^{-2\Delta_1}$ dependence, as opposed to  $t^{-\Delta_1}$ in the \sch\ case \eqref{sch2pt}.
This is of course correct, since the two-point function has scaling dimension $2\Delta_1$, and
 $t$ has dimension one here, while it has dimension  two in the \sch\ case.

\section{Field theories with exotic conformal Galilei symmetry}
\label{free-theory}

\subsection{Free field action}
\label{freeaction}

In Sec.~\ref{secthree} we showed that a field saturating the unitarity bound $\Delta=1$ 
satisfies equation \eqref{MTeqn}, arising also from the non-relativistic contraction of the massless Dirac equation. 
This is the equation of motion of the free field action
\begin{equation}
S=\int \diff t\diff^2\vec x  \pairing{\bar\psi}{(\partial_t + \frac{1}{i\theta}\epsilon_{ij}\chi_i \partial_j)\psi}
\end{equation} 
where $\pairing{\bar\phi}{\psi}$ is an inner product
on the space on which $\chi_i$ act, so that $\chi_i$ is anti-Hermitean.
When we use $v_{1,2}$ to represent $\chi_{1,2}$, 
it is given by the integration over $v_{1,2}$, i.e. 
\begin{equation}
\pairing{\bar\phi}{\psi}=\int \diff^2\vec v \, \bar\phi\psi.
\end{equation} 
We can check that this action is indeed invariant under the exotic
conformal Galilei symmetry, \eqref{action-on-fields} when $\Delta=1$.
Invariance under $H$, $P_i$, $D$ is immediate. 
Invariance under $C$ then implies invariance under $K_i$ and $F_i$,
so let us check that. 
The change of the action is, using integration by parts, 
\begin{multline}
\int \diff t\diff^2\vec x \left[\pairing{\overline{C\psi}}{(\partial_t + \frac{1}{i\theta}\epsilon_{ij}\chi_i \partial_j)\psi} + 
\pairing{\overline{\psi}}{(\partial_t + \frac{1}{i\theta}\epsilon_{ij}\chi_i \partial_j)C\psi}\right] \\
= \int \diff t\diff^2x \left[2t \pairing{\bar\psi}{(\partial_t + \frac{1}{i\theta}\epsilon_{ij}\chi_i \partial_j)\psi}
- \pairing{\bar\psi}{[C,\partial_t + \frac{1}{i\theta}\epsilon_{ij}\chi_i \partial_j]\psi}\right]~.
\label{Cinv}
\end{multline}
Now, an explicit calculation shows \begin{equation}
[C,\partial_t + \frac{1}{i\theta}\epsilon_{ij}\chi_i \partial_j]
= 2t (\partial_t + \frac{1}{i\theta}\epsilon_{ij}\chi_i \partial_j) \label{C-kin-commutator}
\end{equation} therefore we conclude this action is invariant under the exotic conformal symmetry.

When the operators $\chi_{1,2}$ and $\Theta$ are realized using the coordinates $v_{1,2}$ and $\xi$,
the action above becomes
\begin{align}
S&=-\int \diff t \diff \xi \diff^2\vec x \diff^2 \vec v\;( \partial_\xi\bar\psi \partial_t + \epsilon_{ij} \chi_i \bar\psi \partial_j \psi) \nn\\
&= -\int \diff^6 x \sqrt{-g} g^{IJ} \partial_I \bar\psi \partial_J \psi \label{vaction}
\end{align}
 where $I$ runs from $1$ to $6$, corresponding to 
the coordinates $t$, $\xi$, $x_1$, $x_2$, $v_1$, $v_2$,
and the metric is given by 
\begin{equation}
\diff s^2 =  \diff t (\diff \xi- \frac12 \epsilon_{ij}v_i \diff v_j )  
-\epsilon_{ij} \diff x_i \diff v_j ~. 
 \label{R33metric}
\end{equation} This is in fact a metric on flat $\R^{3,3}$. Defining 
\begin{equation}
y_i = x_i +\frac{1}{2} tv_i ~, 
\label{v-to-flat}
\end{equation} 
the metric becomes 
\begin{equation}
\diff s^2 =  \diff t \diff \xi -\epsilon_{ij} \diff y_i \diff v_j ~. \label{flat-metric}
\end{equation}
One finds that the vector fields \eqref{l1conf} acts on this flat metric
as a conformal transformation. Then the invariance of the action \eqref{vaction}
with a suitable weight on $\psi$ is automatic, because a free relativistic scalar field
on a flat space is relativistically conformally invariant.

This is clearly parallel to the fact that the invariance of the 
action of a free non-relativistic massive particle 
\begin{equation}
S=\int \diff t\diff^d x  \bar\psi(i\partial_t + \frac{1}{2m}\partial_i\partial_i)\psi
\end{equation} under the \sch\ symmetry. Indeed, 
by introducing the direction $\xi$ conjugate to $m$, 
the action can be written 
as \begin{equation}
S=-\int \diff t\diff \xi\diff^d x  (2\partial_\xi\bar\psi \partial_t \psi 
+ \partial_i\bar\psi\partial_i\psi)
\end{equation} 
and this describes a free relativistic particle on a flat space $\R^{d+1,1}$
with the metric \eqref{bargmetric}.

\subsection{Chern-Simons-matter action} 

The free field action can be coupled to a $U(1)$ gauge field, together with a Chern-Simons term, without spoiling the Galilei conformal invariance, at least classically. The total action is
\begin{equation}
S=\int \diff t\diff^2\vec x  \left[-\frac\lambda 2\epsilon_{abc}A_a \partial_b A_c  + \pairing{\bar\psi}{(D_t + \frac{1}{i\theta}\epsilon_{ij}\chi_i D_j)\psi}\right],
\end{equation} where $A_a$ is the gauge field and 
 $\lambda$ is the Chern-Simons level.
We use the convention that the indices $a,b,c=0,1,2$ stand for
the directions $t, x_1, x_2$ combined. $D_a$ is the covariant derivative \begin{equation}
D_a=\partial_a + i A_a.
\end{equation} Note that $A_a$ only depends on $t,x_{1,2}$,
and we do not have components $A_{v_i}$ or $A_{\xi}$.
The conformal Galilei group acts on $A_a$ by the Lie derivative $\lie$
via the vector fields \eqref{lcgal:oper}.
Being topological, the Chern-Simons term is automatically invariant under 
the conformal Galilei group. 
We let the exotic conformal Galilei group act on $\psi$ as before, i.e.~{\em without} changing
$\partial_a$ to $D_a$ in \eqref{action-on-fields}.
Invariance of the action under $H$, $P_i$ and $D_i$
is again trivial. The change under $C$ of the Lagrangian is now
\begin{multline}
\int \diff t\diff^2\vec x \Bigl[2t \pairing{\bar\psi}{(D_t + \frac{1}{i\theta}\epsilon_{ij}\chi_i D_j)\psi}\\
- \pairing{\bar\psi}{[C,D_t + \frac{1}{i\theta}\epsilon_{ij}\chi_i D_j]\psi} 
- \pairing{\bar\psi}{[(\lie_C A)_t + \frac{1}{i\theta}\epsilon_{ij}\chi_i (\lie_C A)_j]\psi}\Bigr]
\end{multline} which vanishes using the fact that 
\begin{equation}
[C,D_a] + (\lie_C A)_a = 2t D_a.
\end{equation} 
Therefore the action is invariant under $C$ and invariance under $K_i$ and $F_i$ follows automatically.

The equations of motion for the gauge field are then 
\begin{equation}
F_{12}=  \frac{ i}{\lambda}\pairing{\bar\psi}{\psi}, \qquad
F_{0i}= \frac{i}{\lambda\theta}\pairing{\bar\psi}{\chi_i\psi}~.
\label{CSeom}
\end{equation} 
Let us stress again that  the gauge potential $A_a$ and the field strength $F_{ab}$ only depend on $t,x_{1,2}$.
When we represent $\chi_i$ using $v_{i}$, the pairing in the right hand side of
\eqref{CSeom} involves the integration along the $v_{i}$ directions, e.g. 
\begin{equation}
F_{12}=\frac{i}{\lambda}\int \diff^2\vec v\, \bar\psi \psi~.
\end{equation}
It would be nice to study the coupled CS-matter system in more detail, but we
leave this for future work, and return to discuss the free model.

\subsection{Free field wave function}

We will now construct simple solutions to the free field equation \eqref{MTeqn}
in the representation (\ref{vrep}), analogous to the plane-wave solutions of the \sch\ equation\footnote{Other representations of the Heisenberg generators $\chi_i$
are possible, but we will not consider them here.}. In particular, we will discuss the transformation properties under the exotic conformal symmetry group, showing that the wave functions are projective representations of this group. These solutions 
will be also useful later in the ``holographic'' calculation in Sec. \ref{metricsec}.

The quickest way to arrive at the solution
is to employ a change of coordinates 
which manifestly transforms the operator $\Theta H+\vec{K}\times \vec{P}$ 
to the Laplacian on $\R^{3,3}$ as we used in  sub-section \ref{freeaction}.
Using \eqref{v-to-flat}
the wave equation becomes 
\begin{equation}
\left[\frac{\de}{\de \xi} \frac{\de}{\de t } + \epsilon_{ij} \frac{\de}{\de v_i}\frac{\de}{\de y_j}\right] \psi  =  0~.
\end{equation}
The solutions are spanned by plane waves
\bea
\psi_{E,\vec p,\vec k}  =  \exp i \left( \theta \xi + E t  - \vec{p}\cdot  \vec{y} - \frac12 \vec{k}\cdot  \vec{v} \right), \qquad
E  =  \frac{1}{2\theta} ( \vec p \times \vec k)
\label{spanned}
\eea 
which in the original coordinates  read
\bea
\psi_{E,\vec p,\vec k} = \exp i\Big( \theta \xi + E t - \vec{p}\cdot  \vec{x}  - \frac{1}{2} (\vec{k} +t\vec{p})\cdot \vec{v}  \Big)~.
\label{wavefunction}
\eea
This shows that the parameter $\theta$ behaves as ``mass'' for a \sch\ equation in $\R^{2,2}$.

Let us now address invariance of this solution under the conformal Galilei symmetries.  
The finite boost $K(\vec b)$ transformation  \eqref{l1boost} 
acts on the wavefunction as
\begin{equation}
U_{K(\vec b)} \psi(X) = \psi(K(-\vec b) X)
\end{equation} where $X$ stands for a point in the extended space time,
$X=(t,\xi,\vec x,\vec y)$. 
It acts on the plane waves as 
\begin{equation}
U_{K(\vec b)} \psi_{E,\vec p,\vec k} = \ex^{\frac i2\vec{k}\cdot \vec{b}} \psi_{E-\frac12 \vec p \cdot\vec b,\vec p,k_i+\theta\epsilon_{ij} b_j},
\end{equation} i.e. one has \begin{equation}
p_i \to p_i, \qquad
k_i\to k_i +\theta \epsilon_{ij} b_j,\qquad
E\to E-\frac12 p_i b_i~.
\end{equation}
In particular, we see that while $\vec{p}$ is clearly the momentum associated to the wave function, $\vec{k}$ may be interpreted as the ``boost''. Indeed the transformations of $\vec{k}$ under the full conformal Galilei group are compatible with this interpretation. 
It is easy now to check that \begin{equation}
U_{K(\vec b)}U_{K(\vec b')} =\ex^{i\theta \vec b \times \vec b'} U_{K(\vec b')}U_{K(\vec b)}
\end{equation} with the required phase term coming from the central charge
$[K_i,K_j]=\epsilon_{ij} i\theta$.
Next consider the accelerations. One can take 
\eqref{l1acc} as the finite transformation, which we denote by $F(\vec a)$.
We then let it act on the wavefunction as
\begin{equation}
U_{F(\vec a)} \psi(X) = \psi(F(-\vec a) X).
\end{equation}  
It acts on the plane wave as follows: \begin{equation}
U_{F(\vec a)} \psi_{E,\vec p,\vec k} = \psi_{E- \vec k \cdot\vec a,p_i-2\theta\epsilon_{ij}a_j ,k_i},
\end{equation} i.e. one has \begin{equation}
p_i\to p_i-2\theta\epsilon_{ij}a_j, \qquad
k_i \to k_i ,\qquad
E\to E- k_i  a_i~.
\end{equation}
Recalling 
\begin{equation}
U_{P(\vec c)} \psi_{E,\vec p,\vec k}= \ex^{-i \vec c \cdot\vec p} \psi_{E,\vec p,\vec k},
\end{equation} one can easily see that
\begin{equation}
U_{P(\vec c)} U_{F(\vec a)}= \ex^{-2i\theta \vec c \times \vec a}U_{F(\vec a)}U_{P(\vec c)} ,
\end{equation} thus realizing $[P_i,F_j]=-2\epsilon_{ij} i\theta$.
Let us finally address conformal transformations generated by $C$.
The infinitesimal version is given in \eqref{l1conf}, and
the integrated version is 
\begin{align}
t&\to \frac{t}{1-at}, & x_i&\to  \frac{x_i}{(1-at)^2}, \nn\\
v_i &\to v_i-\frac{2x_i}{t(1-at)}, &
\xi &\to \xi + \frac{2}{t(1-at)} \epsilon_{ij} x_i v_j.
\end{align} Let us denote this action by $C(a)$. This affects the metric \eqref{R33metric} 
as \begin{equation}
\diff s^2 \to  \frac{ \diff s^2 }{(1-at)^2}~.
\end{equation}
Then the action on the wavefunction is \begin{equation}
U_{C(a)} \psi (X) = (1+at)^2\psi(C(-a)X) ~.
\end{equation}
This does not map plane waves to plane waves, but it correctly maps solutions to solutions.

\subsection{Two-point functions}

It is straightforward calculate the two-point functions of the free theory in the $v_i$ representation.
In the six-dimensional flat space, the two-point function of the relativistic 
massless free scalar field $\psi$
is given by \begin{equation}
\langle \psi(x_{(1)}^I) \psi(x_{(2)}^I) \rangle \propto |x_{(1)}^I-x_{(2)}^I|^{-4}.
\end{equation} Using the coordinate change \eqref{v-to-flat} from the $v_i$ representation to the
flat $\R^{3,3}$, one finds \begin{equation}
\langle \psi(x_{(1)}^I) \psi(x_{(2)}^I) \rangle \propto (t\xi-\epsilon_{ij}(x_i+\frac{t v_{+j}}4 )v_j   )^{-2}.
\end{equation} In order to compare this to the results in Sec.~\ref{2pt-func},
one needs to perform the Fourier transformation along $\xi$. 
Then  the two-point function is  
\begin{equation}
\int \diff \xi e^{i\theta\xi} (t\xi-\epsilon_{ij}(x_i+\frac{t v_{+j}}4 )v_j   )^{-2}
= t^{-2} \theta \exp( i\theta \epsilon_{ij} (u_i+\frac{ v_{+j}}4 )v_j  ) \int\diff\xi e^{i\xi} \xi^{-2}.
\end{equation}
The last term is just an numerical coefficient. Indeed, this is exactly of the form \eqref{2pt-func-final}, where $g(\vec u + \vec v_{+}/2) \equiv 1$ and $\Delta=1$.

\section{Geometric realization}
\label{metricsec}

\subsection{Invariant metric}

In the previous section we saw that 
one way to realize the free field with the exotic conformal Galilei invariance
is to use a free ``relativistic'' scalar field on flat $\R^{3,3}$.
The exotic conformal Galilei invariance acts on $\R^{3,3}$
as a relativistic conformal transformation. Then it is easy to add another direction
to realize the whole symmetry as the isometry. Indeed, by taking the seven-dimensional
metric 
\begin{equation}
g = \frac{1}{z^2}\left[ \diff z^2-
 \diff t (\diff \xi - \frac{1}{2} \epsilon_{ij} v_i\diff v_j )   +\epsilon_{ij} \diff x_i \diff v_j  
\right]
\end{equation} 
which is just a rewriting of ``AdS$_7$'' with $(3,4)$ signature,
we can realize the exotic conformal Galilei algebra using isometries. 
The only generators modified by terms in 
$z$ are the dilatations and the conformal transformations. In particular, we have
\bea
D= - x_i \frac\partial{\partial x_i} - t \partial_t - \frac12z\partial_z,  \quad  
C= 2 t x_i \frac\partial{\partial x_i} - 2x_i \frac\partial{\partial v_i} + t^2 \partial_t + (z^2+\epsilon_{ij} x_i v_j) \partial_\xi + tz \partial_z~.
\eea
By analyzing general symmetric two-tensors invariant under the transformations 
\eqref{htra} -- \eqref{ttra},   we find a three-parameter family of metrics, 
invariant under this set of isometries, namely:
 \begin{equation}
g = \frac{a}{z^2}\left[ \diff z^2
  - \diff t (\diff \xi - \frac{1}{2} \epsilon_{ij} v_i\diff v_j)+\epsilon_{ij} \diff x_i \diff v_j  
\right] + b\frac{\diff t^2}{z^4}+ \frac{c}{z^4}(\diff x_i + v_i \diff t)^2~.
\label{finalmetric}
\end{equation}
Unfortunately, it turns out that the signature of this metric is
always $(3,4)$.  For this reason, we can not interpret this as bulk metric of a consistent 
gravity theory in seven dimensions.

Let us write down the Laplacian acting on scalars, constructed 
from the metric (\ref{finalmetric}): 
\bea
 && \Box = \frac{z^7}{a}\frac{\de}{\de z}\left( \frac{1}{z^5}\frac{\de}{\de z}\right) - \frac{4z^2}{a}\left(\frac{\de}{\de t} \frac{\de}{\de \xi} + \epsilon_{ij}\chi_i \frac{\de}{\de x_j}\right)  -\frac{4b}{a^2}\frac{\de^2}{\de \xi^2}\nn\\[2mm]
&&\qquad \qquad \qquad  \qquad \qquad \qquad 
- \frac{4c}{a^2} \left(\chi_i \chi_i - 2 \epsilon_{ij}v_i \frac{\de}{\de v_j}\frac{\de}{\de \xi}\right)
\eea
Notice that the second and third terms  in the first line are the quadratic 
Casimir operators
$H\Theta + \vec{K}\times \vec{P}$ and $\Theta^2$ of the exotic Galilei group (without conformal extension).

\subsection{``Holographic'' two-point function}

Although the metric \eqref{finalmetric} has the wrong signature, recall that setting 
$b=c=0$, this  is an analytic continuation of AdS$_7$. Furthermore, 
when $c=0$, this metric is an analytic continuation of the metric which has
the \sch\ group as the isometry, constructed in \cite{son,mcgreevy}. 
It is therefore natural to wonder whether applying 
the AdS/CFT rules to this metric might still give sensible results, in spite of the problematic signature. 
In the following, we then ignore this problem, and perform 
a holographic-type calculation of the two-point function. Holographic two- and three- point functions in \sch\ backgrounds have been recently computed in \cite{Volovich:2009yh,Fuertes:2009ex}. However, our treatment is more elementary and follows essentially the computation of two-point functions in \cite{mcgreevy}.

We then imagine a massive scalar field propagating in the fixed background metric \eqref{finalmetric}, 
and solve for its wave equation with fixed boundary conditions, following the original prescription \cite{Gubser:1998bc,Witten:1998qj}. As we will see, as far as this naive computation is concerned, 
the wrong signature of the boundary metric does not play an important role. 

For simplicity, we consider the metric \eqref{finalmetric}  after 
setting\footnote{Solutions in the metric with $c\neq 0$ can also be studied, but they can not be expanded into the free wave functions as we do below.}  $c=0$. 
We can  set $a=1$ by an overall rescaling of the metric. The parameter $b$ may also be reabsorbed, but we 
leave it explicit.  Using the ansatz
\bea
\phi = \psi_\mathrm{free}(t,\vec{x},\xi,\vec{v}) \, \varphi(z)
\eea
into the Klein-Gordon equation, we get
\bea
-z^7\frac{\de}{\de z}\left( \frac{1}{z^5}\frac{\de}{\de z}\right) \varphi (z) + \kappa^2 z^2 \varphi(z) +(m^2 - 4b \theta^2) \varphi(z) = 0 
\label{holowave}
\eea
where we defined
\bea
\kappa^2 \equiv - 4 (E \theta + \frac{1}{2} \vec{k}\times \vec{p})~.
\eea
In particular, we require that $\kappa^2>0$. Defining 
\bea
\nu = \sqrt{9 + m^2 - 4b \theta^2}, \qquad \quad \hat{\Delta} = 3 + \nu
\label{delta}
\eea
the solution to \eqref{holowave}, normalizable near the boundary $z\to 0$ is given by the Bessel function 
\bea
\varphi (z) = C z^3 K_\nu (\kappa z)~,
\label{bessel}
\eea
where $C$ is a constant. In particular, $\varphi\sim z^{\hat{\Delta}}$ for $z \to 0$. The reason why we decorate $\Delta$ with a hat will become clear momentarily. Notice that this solution is formally the same solution 
for the Klein-Gordon equation in a Sch$_4$ background \cite{mcgreevy}.

From the asymptotic expansion of (\ref{bessel}) one extracts the ``flux factor'' at the boundary, which is essentially the Fourier transform of the two-point function in momentum space. For us ``momentum space'', 
means the space of $p=(E,\vec{p},\theta,\vec{k})$. Thus the two-point function is 
\bea
\langle \cO_1 (t_1,\vec{x}_1,\xi_1,\vec{v}_1)\cO_2 (t_2,\vec{x}_2,\xi_2,\vec{v}_2)\rangle = 
\int \prod_{i=1}^2
 \diff \theta_i \diff E_i \diff^2\vec{p}_i \diff^2 \vec{k}_i  
\langle {\cal O}_1(p_1){\cal O}_2(p_2) \rangle
\psi_{\mathrm{free}_1} \psi_{\mathrm{free}_2} \nonumber \\
\eea
where $\langle {\cal O}_1(p_1){\cal O}_2(p_2) \rangle$
is the momentum-space two-point function, and 
$\psi_\mathrm{free}$ are the solutions \eqref{wavefunction}.
Making the change of variables
$\vec{x} \to \vec{y} = \vec{x} + t\vec{v}/2$
the free wave functions become simply the plane waves \eqref{spanned}.
It is then clear that 
solutions to the wave equation can be expanded into these plane waves as 
\bea
\phi (t,\vec{y},\xi,\vec{v}) = \int\diff E \diff\theta \diff^2 \vec{p} \diff^2 \vec{k} \,\hat \phi_p \, \psi_\mathrm{free}
\eea
and we have
\bea
\langle {\cal O}_1(p_1){\cal O}_2(p_2) \rangle = \delta(\theta_1+\theta_2)\delta(E_1+E_2)\delta^2(\vec{p}_1+\vec{p}_2)\delta^2(\vec{k}_1+\vec{k}_2) \langle \dots \rangle  ~,
\eea
where $\langle \dots \rangle$ is extracted from the flux factor.  
Changing coordinates as in \eqref{newvar}
we can carry out the integral over the delta-functions, reducing it to 
\bea
\int \diff\theta_1  \diff E_1  \diff\vec{p}_1  \diff\vec{k}_1 \langle \dots \rangle 
\ex^{ i \left[ \theta_1 \xi  + E_1 t   - \vec{p}_1  \cdot \vec{y} -  \frac{1}{2}\vec{k}_1 \cdot \vec{v}\right]}~,
\label{integ}
\eea
where notice that
\bea
\vec{y} = \vec{x} + \frac{1}{4}(  t \vec{v}_+ + t_+ \vec{v})~. 
\eea
We will now evaluate \eqref{integ}, where $\langle \dots \rangle\propto \kappa^{2\nu}$,
without being concerned with the overall numerical coefficient. 
After the following change of variables 
\bea
 z = t (E_1 + \frac{1}{2\theta_1} \vec{k}_1\times {\vec p}_1 )\qquad \vec{p}_1 = \vec{p}\theta t^{-1/2} \qquad \vec{k}_1 = \vec{k}\theta t^{-1/2}
\eea 
we have
\bea
\Gamma (\nu+1) t^{-\hat{\Delta}} \int \diff \theta_1 \theta_1^{\hat{\Delta}+1}\ex^{i\theta_1 \xi}\int \diff \vec{p}  \diff \vec{k}\exp \left[-i \theta_1\left(\frac{1}{2} \vec{k}  \times \vec{p} + \frac{\vec{p}\cdot \vec{y}}{ t^{1/2}}
 +\frac{\vec{k} \cdot \vec{v}}{2 t^{1/2}}\right)\right]~.
\eea
The latter integral is easily evaluated by analytically continuing $i\theta_1 \to \theta_1$,  making it just a Gaussian integral.
In conclusion, we  obtain 
\bea
\langle \cO_1 (t_1,\vec{x}_1,\xi_1,\vec{v}_1)\cO_2 (t_2,\vec{x}_2,\xi_2,\vec{v}_2)\rangle \propto t^{-\hat{\Delta}} \int \diff \theta_1 \theta_1^{\hat{\Delta}-1}\ex^{i\theta_1 \xi} \exp \left[i\theta_1 \vec{v}\times 
\left(\frac{\vec{x}}{t}+ \frac{1}{4}\vec{v}_+\right) \right]~.
\eea
where notice that the dependence on $t_1+t_2$ through $\vec{y}$ dropped out of the final expression, as 
expected from invariance under time translations. We see that we get agreement with the general expression 
 \eqref{2pt-func-final}, provided we identify 
\bea
\hat \Delta = 2 \Delta_1~.
\eea
Indeed, notice that lower bound on $\hat\Delta$ is $\hat\Delta_\mathrm{min}=2$. This corresponds to the unitarity bound for the free \sch\ theory in $d=4$ dimensions, or equivalently for a relativistic CFT in $D=5+1$ dimensions.
As an exotic   Galilean conformal field theory in $d=2$ dimension, it correctly corresponds to the unitarity bound $\Delta_\mathrm{min}=1$.  The factor of $\theta_1^{2\Delta-1}$ is analogous to the  factor of $M_1^{\Delta-1}$ appearing in the holographic two-point function for \sch\ theories \cite{Volovich:2009yh}.

\section{Conclusions}
\label{conclusions}

In this paper we have discussed non-relativistic
conformal  systems with symmetries different from the \sch\ type, and 
their geometric realizations. 
We started by analyzing a family of conformal Galilei algebras 
which are natural generalizations of the \sch\ algebra. In particular, we have pointed out the existence of central extensions and infinite dimensional extensions.
Our main interest was the Galilean algebra extended by constant accelerations and  conformal transformations. This 
 admits a central extension in $d=2$ spatial dimensions. Following analogous discussions of  the \sch\ algebra, 
we have studied generic properties of  
conformal field theories, with underlying conformal Galilei algebra. For example, we have defined
primaries and discussed the general form of two-point functions. Higher $n$-point functions may be analyzed 
in a similar way. As the \sch\ conformal field theories are relevant for studying cold atoms, our 
results should be useful for studying other, perhaps exotic, Galilean conformal field theories describing 
real physical systems \cite{Horvathy:2006gi}.

One crucial ingredient in our approach is the central extension of the algebra. Such extension renders
the boosts along the two directions in the plane non-commutative. Following \cite{Jackiw:2000tz}, one way to understand 
the origin of this non-commutativity is to derive the algebra from a non-relativistic contraction of the relativistic conformal algebra, in a representation with non-zero spin. Although the contraction applied to an AdS metric
is degenerate, we noticed that applying  the  modified contraction  above to the (massless)  Dirac 
equation leads a consistent non-trivial equation, also obeyed by fields whose scaling dimension
saturates the unitarity bound. We have discussed that this equation is
to the  exotic conformal Galilei  group what  the free \sch\ equation is to the \sch\ group.
Moreover, we have shown that one can couple it to a gauge field, together with a Chern-Simons term. This suggests that the contraction we have discussed may be applied to more complicated models, perhaps supersymmetric, like the model of \cite{Aharony:2008ug}. It would be nice to see whether new interesting non-relativistic models
may be obtained in this way.  

Following the strategy that successfully lead to the \sch\ geometries in \cite{son,mcgreevy}, we have presented
a three-parameter metric that realizes the exotic Galilean symmetry group as its group of isometries.
 However, it turns out that this metric has not Lorentzian signature, but rather it has $(3,4)$ signature. Despite this, the geometric structure may be interpreted formally 
in AdS/CFT language, and leads to a modification of the ``AdS$_7$'' metric, where the boundary is $\R^{3,3}$.
The three extra directions $(\xi, \vec{v})$ are associated to the central extension $\Theta$,
and to  \emph{velocities}, respectively. The symmetries act on this space, in a slightly unusual way. For example, dilatations act as $D: (t,\vec{x},\xi,\vec{v}) \to (\lambda t,\lambda \vec{x},\xi,\vec{v})$.

 Although this is a 
natural geometric (Bargmann-like)  structure generalizing the \sch\ case, the resulting metric is not suitable for AdS/CFT applications. Nevertheless, we have pointed out that  a blind application of the AdS/CFT 
prescriptions gives a consistent ``holographic'' two-point function. We suspect  that this feature 
will persist for three- and higher point functions. 
 It would be interesting to see whether the geometric realization that we have presented, may be still useful at least as a  tool for performing other holographic-type calculations.

The problem of finding a gravity dual of non-\sch\ 
Galilean conformal field theories remains largely open. 
From our analysis, it emerged that the inclusion of fermionic variables in taking the non-relativistic limit
plays an important role.  Thus, one can try  to obtain a bulk dual by applying 
the non-relativistic contraction to  the matter-coupled supergravity equations, presumably 
including fermionic fields. 
A different approach to a geometric realization of the
conformal Galilei algebra (without central extension) 
has been proposed in \cite{Bagchi:2009my}. In this reference it is argued  that
the gravity side should be described by a version of Newton-Cartan theory, where the dynamical variables
are non-metric connections. However, the relation of this approach to the AdS/CFT correspondence remains to be clarified. It should be interesting  to investigate the relationship of our results to those of   \cite{Bagchi:2009my}.

\section*{Acknowledgments}

YT is supported in part by the NSF grant PHY-0503584,
and in part by the Marvin L. Goldberger membership in the Institute
for Advanced Study.
%%%%%%%%%%%%%%%%%%%%%%%%%%%%%%%%%%%%%%%%%%%%%%%%%%%%%%%%%%%%%%%%%%%%%%%%%%%%%%%%%%%%%%%%%%%%%%%%%%%%

\appendix

\section{Mechanics with exotic conformal Galilei symmetry}
\label{lagappe}

The Lagrangian
\bea
L = \frac12m\dot x_i^2  -\theta \epsilon_{ij} \dot x_i \ddot x_j
\eea
is known to be invariant under the Galilei group with two central extensions $M$ and $\Theta$
\cite{Lukierski:1996br}. If $\theta=0$, it is quasi-invariant under the \sch\ group, meaning that it is invariant up to a total derivative. 
It is quasi-invariant under the exotic conformal Galilei group only if $m=0$,
when the Lagrangian is \begin{equation}
L =  -\theta \epsilon_{ij} \dot x_i \ddot x_j\label{seclag}
\end{equation}
whose equation of motion is
\bea
\dddot x_i = 0.
\eea
We see that the solutions are constantly accelerating trajectories, which is analogous to 
the constant velocity trajectories solutions to $L=\frac{m}{2}\dot x_i^2 $. 
Because of the high number of derivatives it is not easy to quantize it directly.
Let us instead consider the Lagrangian
\bea
L = - \theta \left( \epsilon_{ij} \dot x_i \dot v_j + \frac{1}{2} \epsilon_{ij}v_i \dot v_j  \right)~. \label{lag-in-v}
\eea
The equations of motion of this  are 
\bea
\dot v_i = - \ddot x_i, \qquad \ddot v_i=0 ~,
\eea 
thus we see that this Lagrangian is equivalent to \eqref{seclag}, and $v_i$ are essentially (minus) 
the velocities.

One way to rewrite this Lagrangian is to use the metric \eqref{R33metric}.
Let $x^I$ denote $(t,\xi,x_1,x_2,v_1,v_2)$ collectively, and consider 
a free, massless ``relativistic'' 
particle moving in $\R^{3,3}$, with the standard action \begin{equation}
\int \diff\tau\,  e^{-1}(\tau) g_{IJ}(x) \frac{\diff {x^I}}{\diff \tau}\frac{\diff {x^J}}{\diff \tau}\label{relpart}
\end{equation} 
where $\tau$ is the parameter of the worldline $x^I(\tau)$, and
$e(\tau)$ is the einbein.
This is invariant under the relativistic conformal group acting on $x^I$,
which includes the exotic conformal Galilei group,
because the conformal factor can be reabsorbed by the einbein.
Now let us gauge-fix the worldline by choosing  $\tau=t$, which gives the action
\begin{equation}
-\int \diff t\, e^{-1}\left[
-\dot\xi +  \frac12 \epsilon_{ij}v_i \dot v_j 
+\epsilon_{ij} \dot x_i \dot v_j 
\right].\label{A7}
\end{equation} 
Taking the variation with respect to $\xi$ one finds the einbein $e^{-1}$ is constant
which we denote by $\theta$, resulting in 
\begin{equation}
-\theta\int \diff t\, \left[
-\dot\xi +  \frac12 \epsilon_{ij}v_i \dot v_j 
+\epsilon_{ij} \dot x_i \dot v_j 
\right].
\end{equation} 
which is equivalent to \eqref{lag-in-v} up to an apparently decoupled 
degree of freedom $\xi$; more precisely, the equation of motion associated to $e$
in \eqref{A7} determines the time evolution of $\xi$ in terms of that of $x_i$ and $v_i$.
One can check that 
the Lagrangian \eqref{lag-in-v} without $\xi$ is invariant under the conformal
Galilei transformation up to a total derivative, which is precisely absorbed
by a shift of $\xi$ given in \eqref{xi-shift}.

For comparison, it  is instructive to carry out the same analysis of the conformal symmetry
to the case of free non-relativistic particle with mass $m$.
Let us consider again the Lagrangian of a massless relativistic particle \eqref{relpart},
this time with the Bargmann metric \eqref{bargmetric} as $g_{IJ}$.
 Gauge-fixing by $\tau=t$, one obtains the Lagrangian
\begin{equation}
m\int \diff t\,\left[\dot\xi+\frac 12 \dot x_i \dot x_i\right].
\end{equation}  
which is the Lagrangian of a non-relativistic massive particle.
This is invariant under the \sch\ transformations \eqref{bargs}. Equivalently,
it shows the invariance up to total derivative under the \sch\ 
symmetry of the Lagrangian $L=\tfrac{m}{2}\dot x_i^2$.

Let us construct the Hamiltonian for the Lagrangian \eqref{lag-in-v}.
We calculate the conjugate momentum for $x_i$ and $v_i$ via
\bea
p_i = \frac{\de L}{\de \dot x_i}=-\theta\epsilon_{ij} \dot v_j,
\qquad \quad \pi_i = \frac{\de L}{\de \dot v_i}= \theta\epsilon_{ij}(\dot x_j+\frac12 v_j) ~
\eea 
and find
\bea
H = p_i \dot x_i + \pi_i \dot v_i - L = 
- \frac{1}{\theta}\epsilon_{ij}\left(\pi_j - \frac{\theta}{2}\epsilon_{jk}v_k \right)p_j~.
\eea 
Quantization proceeds by the replacement \begin{equation}
p_i \to -i\frac{\partial}{\partial x_i}, \qquad
\pi_i \to -i\frac{\partial}{\partial v_i}. 
\end{equation} Then the corresponding wave equation is \begin{equation}
\frac{\partial}{\partial t}\psi = 
-iH\psi = -
\frac{1}{i\theta}\epsilon_{ij}(\frac{\partial}{\partial v_i} -\frac{i\theta}2\epsilon_{jk}v_k  )
\frac{\partial}{\partial x_j}\psi~.
\end{equation}
This is exactly the free wave equation \eqref{MTeqn} discussed in the main text.


\begin{thebibliography}{99}

\bibitem{son}

  D.~T.~Son,
  ``Toward an AdS/Cold Atoms Correspondence: a Geometric Realization of the   Schr\"odinger Symmetry,''
{\slshape   Phys.\ Rev.\  D }{\bf 78} (2008) 046003
  [arXiv:0804.3972 [hep-th]].
  
%%CITATION = PHRVA,D78,046003;%%
\bibitem{mcgreevy}

  K.~Balasubramanian and J.~McGreevy,
  ``Gravity Duals for Non-Relativistic CFTs,''
{\slshape   Phys.\ Rev.\ Lett.\  }{\bf 101} (2008) 061601
  [arXiv:0804.4053 [hep-th]].
  
%%CITATION = PRLTA,101,061601;%%
\bibitem{Nishida:2007pj}

  Y.~Nishida and D.~T.~Son,
  ``Nonrelativistic Conformal Field Theories,''
{\slshape   Phys.\ Rev.\  D }{\bf 76} (2007) 086004
  [arXiv:0706.3746 [hep-th]].
  
%%CITATION = PHRVA,D76,086004;%%
\bibitem{Maldacena:2008wh}

  J.~Maldacena, D.~Martelli and Y.~Tachikawa,
  ``Comments on String Theory Backgrounds with Non-Relativistic Conformal   Symmetry,''
{\slshape   JHEP }{\bf 0810} (2008) 072
  [arXiv:0807.1100 [hep-th]].
  
%%CITATION = JHEPA,0810,072;%%
\bibitem{Herzog:2008wg}

  C.~P.~Herzog, M.~Rangamani and S.~F.~Ross,
  ``Heating Up Galilean Holography,''
{\slshape   JHEP }{\bf 0811} (2008) 080
  [arXiv:0807.1099 [hep-th]].
  
%%CITATION = JHEPA,0811,080;%%
\bibitem{Adams:2008wt}

  A.~Adams, K.~Balasubramanian and J.~McGreevy,
  ``Hot Spacetimes for Cold Atoms,''
{\slshape   JHEP }{\bf 0811} (2008) 059
  [arXiv:0807.1111 [hep-th]].
  
%%CITATION = JHEPA,0811,059;%%
\bibitem{BurdetPerrinSorba}
G.~Burdet, M.~Perrin and P.~Sorba, ``About the Non-Relativistic Structure of the Conformal Algebra,''{\slshape  Commun. Math. Phys. }{\bf 34} (1973) 85.
\bibitem{Duval:1984cj}

  C.~Duval, G.~Burdet, H.~P.~Kunzle and M.~Perrin,
  ``Bargmann Structures and Newton-Cartan Theory,''
{\slshape   Phys.\ Rev.\  D }{\bf 31} (1985) 1841.
  
%%CITATION = PHRVA,D31,1841;%%
\bibitem{Duval:1990hj}

  C.~Duval, G.~W.~Gibbons and P.~Horv\'athy,
  ``Celestial Mechanics, Conformal Structures, and Gravitational Waves,''
{\slshape   Phys.\ Rev.\  D }{\bf 43} (1991) 3907
  [arXiv:hep-th/0512188].
  
%%CITATION = PHRVA,D43,3907;%%
\bibitem{Bergman:2000cw}

  A.~Bergman and O.~J.~Ganor,
  ``Dipoles, Twists and Noncommutative Gauge Theory,''
{\slshape   JHEP }{\bf 0010} (2000) 018
  [arXiv:hep-th/0008030].
  
%%CITATION = JHEPA,0010,018;%%
\bibitem{Kachru:2008yh}

  S.~Kachru, X.~Liu and M.~Mulligan,
  ``Gravity Duals of Lifshitz-Like Fixed Points,''
{\slshape   Phys.\ Rev.\  D }{\bf 78} (2008) 106005
  [arXiv:0808.1725 [hep-th]].
  
%%CITATION = PHRVA,D78,106005;%%
\bibitem{negro1}
J.~Negro, M.~A.~del Olmo and A.~Rodr\'iguez-Marco,  ``Nonrelativistic Conformal Groups. I,''{\slshape   J. Math. Phys. }{\bf 38} (1997) 3786.

\bibitem{negro2}
J.~Negro, M.~A.~del Olmo and A.~Rodr\'iguez-Marco,  ``Nonrelativistic Conformal Groups. II. Further Developments and Physical Applications,''{\slshape   J. Math. Phys. }{\bf 38} (1997) 3810.

\bibitem{Duval:2000xr}
  C.~Duval and P.~A.~Horv\'athy,
  ``The exotic Galilei group and the `Peierls substitution',''
  {\slshape Phys.\ Lett.\  B} {\bf 479}, 284 (2000)
  [arXiv:hep-th/0002233].
  %%CITATION = PHLTA,B479,284;%%



\bibitem{Bhattacharyya:2008kq}

  S.~Bhattacharyya, S.~Minwalla and S.~R.~Wadia,
  ``The Incompressible Non-Relativistic Navier-Stokes Equation from Gravity,''
  arXiv:0810.1545 [hep-th].
  
%%CITATION = ARXIV:0810.1545;%%
\bibitem{Fouxon:2008ik}

  I.~Fouxon and Y.~Oz,
  ``CFT Hydrodynamics: Symmetries, Exact Solutions and Gravity,''
{\slshape   JHEP }{\bf 0903} (2009) 120
  [arXiv:0812.1266 [hep-th]].
  
%%CITATION = JHEPA,0903,120;%%
\bibitem{Horvathy:2006gi}

  P.~A.~Horv\'athy,
  ``Non-Commutative Mechanics, in Mathematical \& in Condensed Matter Physics,''
  arXiv:cond-mat/0609571.
  
%%CITATION = COND-MAT/0609571;%%

\bibitem{Stichel:2003kh}

  P.~C.~Stichel and W.~J.~Zakrzewski,
  ``A New Type of Conformal Dynamics,''
{\slshape   Annals Phys.\  }{\bf 310} (2004) 158
  [arXiv:hep-th/0309038].
  
%%CITATION = APNYA,310,158;%%
\bibitem{Lukierski:2007nh}

  J.~Lukierski, P.~C.~Stichel and W.~J.~Zakrzewski,
  ``Acceleration-Extended Galilean Symmetries with Central Charges and Their   Dynamical Realizations,''
{\slshape   Phys.\ Lett.\  B }{\bf 650} (2007) 203
  [arXiv:hep-th/0702179].
  
%%CITATION = PHLTA,B650,203;%%


\bibitem{Duval:2008jg}

  C.~Duval, M.~Hassaine and P.~A.~Horv\'athy,
  ``The Geometry of Schr\"odinger Symmetry in Gravity   Background/Non-Relativistic CFT,''
{\slshape   Annals Phys.\  }{\bf 324} (2009) 1158
  [arXiv:0809.3128 [hep-th]].
  
%%CITATION = ARXIV:0809.3128;%%
\bibitem{Bargmann:1954gh}

  V.~Bargmann,
  ``On Unitary Ray Representations of Continuous Groups,''
{\slshape   Annals Math.\  }{\bf 59} (1954) 1.
  
%%CITATION = ANMAA,59,1;%%
\bibitem{Inonu:1953sp}

  E.~\.{I}n\"on\"u and E.~P.~Wigner,
  ``On the Contraction of Groups and Their Represenations,''
{\slshape   Proc.\ Nat.\ Acad.\ Sci.\  }{\bf 39} (1953) 510.
  
%%CITATION = PNASA,39,510;%%
\bibitem{Jackiw:2000tz}

  R.~Jackiw and V.~P.~Nair,
  ``Anyon Spin and the Exotic Central Extension of the Planar Galilei  Group,''
{\slshape   Phys.\ Lett.\  B }{\bf 480} (2000) 237
  [arXiv:hep-th/0003130].
  
%%CITATION = PHLTA,B480,237;%%
\bibitem{Jackiw:2002he}

  R.~Jackiw and V.~P.~Nair,
  ``Remarks on the Exotic Central Extension of the Planar Galilei Group,''
{\slshape   Phys.\ Lett.\  B }{\bf 551} (2003) 166
  [arXiv:hep-th/0211119].
  
%%CITATION = PHLTA,B551,166;%%
\bibitem{Duval:2002cw}

  C.~Duval and P.~A.~Horv\'athy,
  ``Spin and Exotic Galilean Symmetry,''
{\slshape   Phys.\ Lett.\  B }{\bf 547} (2002) 306
  [{\slshape Erratum-ibid.\  B }{\bf 588} (2004) 228]
  [arXiv:hep-th/0209166].
  
%%CITATION = PHLTA,B547,306;%%
\bibitem{Lukierski:2005xy}

  J.~Lukierski, P.~C.~Stichel and W.~J.~Zakrzewski,
  ``Exotic Galilean Conformal Symmetry and Its Dynamical Realisations,''
{\slshape   Phys.\ Lett.\  A }{\bf 357} (2006) 1
  [arXiv:hep-th/0511259].
  
%%CITATION = PHLTA,A357,1;%%

\bibitem{Berenstein:2002jq}
  D.~E.~Berenstein, J.~M.~Maldacena and H.~S.~Nastase,
  ``Strings in Flat Space and PP Waves from ${\mathcal{N}}\!=4$ Super Yang Mills,''
{\slshape   JHEP }{\bf 0204} (2002) 013
  [arXiv:hep-th/0202021].
  
%%CITATION = JHEPA,0204,013;%%



\bibitem{Kruczenski:2008bs}

  M.~Kruczenski and A.~A.~Tseytlin,
  ``Spiky Strings, Light-Like Wilson Loops and PP-Wave Anomaly,''
{\slshape   Phys.\ Rev.\  D }{\bf 77} (2008) 126005
  [arXiv:0802.2039 [hep-th]].
  
%%CITATION = PHRVA,D77,126005;%%
\bibitem{Horvathy:2004fw}

  P.~A.~Horv\'athy and M.~S.~Plyushchay,
  ``Anyon Wave Equations and the Noncommutative Plane,''
{\slshape   Phys.\ Lett.\  B }{\bf 595} (2004) 547
  [arXiv:hep-th/0404137].
  
%%CITATION = PHLTA,B595,547;%%
\bibitem{Bagchi:2009my}

  A.~Bagchi and R.~Gopakumar,
  ``Galilean Conformal Algebras and AdS/CFT,''
  arXiv:0902.1385 [hep-th].
  
%%CITATION = ARXIV:0902.1385;%%
\bibitem{Alishahiha:2009np}

  M.~Alishahiha, A.~Davody and A.~Vahedi,
  ``On AdS/CFT of Galilean Conformal Field Theories,''
  arXiv:0903.3953 [hep-th].
  
%%CITATION = ARXIV:0903.3953;%%
\bibitem{Bagchi:2009ca}

  A.~Bagchi and I.~Mandal,
  ``On Representations and Correlation Functions of Galilean Conformal   Algebras,''
  arXiv:0903.4524 [hep-th].
  
%%CITATION = ARXIV:0903.4524;%%
\bibitem{Henkel:1993sg}

  M.~Henkel,
  ``Schr\"odinger Invariance in Strongly Anisotropic Critical Systems,''
{\slshape   J.\ Statist.\ Phys.\  }{\bf 75} (1994) 1023
  [arXiv:hep-th/9310081].
  
%%CITATION = JSTPB,75,1023;%%
\bibitem{Henkel:2003pu}

  M.~Henkel and J.~Unterberger,
  ``Schr\"odinger Invariance and Space-Time Symmetries,''
{\slshape   Nucl.\ Phys.\  B }{\bf 660} (2003) 407
  [arXiv:hep-th/0302187].
  
%%CITATION = NUPHA,B660,407;%%
\bibitem{Alishahiha:2009nm}

  M.~Alishahiha, R.~Fareghbal, A.~E.~Mosaffa and S.~Rouhani,
  ``Asymptotic Symmetry of Geometries with Schr\"odinger Isometry,''
  arXiv:0902.3916 [hep-th].
  
%%CITATION = ARXIV:0902.3916;%%
\bibitem{Brown:1986nw}

  J.~D.~Brown and M.~Henneaux,
  ``Central Charges in the Canonical Realization of Asymptotic Symmetries: an   Example from Three-Dimensional Gravity,''
{\slshape   Commun.\ Math.\ Phys.\  }{\bf 104} (1986) 207.
  
%%CITATION = CMPHA,104,207;%%
\bibitem{LevyLeblond}
J.~M.~L\'evy-Leblond, ``Galilei Group and Galilean Invariance,'' pp. 221--299 in ``Group Theory and Its Applications, Vol. 2'', (Ed. E. Loebl) Academic Press, New York, 1971.
\bibitem{Grigore:1993fz}

  D.~R.~Grigore,
  ``The Projective Unitary Irreducible Representations of the Galilei Group in   (1+2)-Dimensions,''
{\slshape   J.\ Math.\ Phys.\  }{\bf 37} (1996) 460
  [arXiv:hep-th/9312048].
  
%%CITATION = JMAPA,37,460;%%
\bibitem{Bose:1994sj}

  S.~K.~Bose,
  ``The Galilean Group in (2+1) Space-Times and Its Central Extension,''
{\slshape   Commun.\ Math.\ Phys.\  }{\bf 169} (1995) 385.
  
%%CITATION = CMPHA,169,385;%%
\bibitem{Lukierski:1996br}

  J.~Lukierski, P.~C.~Stichel and W.~J.~Zakrzewski,
  ``Galilean-Invariant (2+1)-Dimensional Models with a Chern-Simons-Like  Term   And D = 2 Noncommutative Geometry,''
{\slshape   Annals Phys.\  }{\bf 260} (1997) 224
  [arXiv:hep-th/9612017].
  
%%CITATION = APNYA,260,224;%%
\bibitem{delOlmo:2005md}

  M.~A.~del Olmo and M.~S.~Plyushchay,
  ``Electric Chern-Simons Term, Enlarged Exotic Galilei Symmetry and   Noncommutative Plane,''
{\slshape   Annals Phys.\  }{\bf 321} (2006) 2830
  [arXiv:hep-th/0508020].
  
%%CITATION = APNYA,321,2830;%%
\bibitem{LLeq}
J.-M.~L\'evy-Leblond, ``Nonrelativistic Particles and Wave Equations,'' \textsl{{\slshape Comm. Math. Phys.  }{\bf 6}, }\textbf{4} (1967), 286-311. 
\bibitem{Nakayama:2008qm}

  Y.~Nakayama,
  ``Index for Non-Relativistic Superconformal Field Theories,''
{\slshape   JHEP }{\bf 0810} (2008) 083
  [arXiv:0807.3344 [hep-th]].
  
%%CITATION = JHEPA,0810,083;%%
\bibitem{Lee:2009mm}

  K.~M.~Lee, S.~Lee and S.~Lee,
  ``Nonrelativistic Superconformal M2-Brane Theory,''
  arXiv:0902.3857 [hep-th].
  
%%CITATION = ARXIV:0902.3857;%%
\bibitem{Volovich:2009yh}

  A.~Volovich and C.~Wen,
  ``Correlation Functions in Non-Relativistic Holography,''
  arXiv:0903.2455 [hep-th].
  
%%CITATION = ARXIV:0903.2455;%%
\bibitem{Fuertes:2009ex}

  C.~A.~Fuertes and S.~Moroz,
  ``Correlation Functions in the Non-Relativistic AdS/CFT Correspondence,''
  arXiv:0903.1844 [hep-th].
  
%%CITATION = ARXIV:0903.1844;%%
\bibitem{Gubser:1998bc}

  S.~S.~Gubser, I.~R.~Klebanov and A.~M.~Polyakov,
  ``Gauge Theory Correlators from Non-Critical String Theory,''
{\slshape   Phys.\ Lett.\  B }{\bf 428} (1998) 105
  [arXiv:hep-th/9802109].
  
%%CITATION = PHLTA,B428,105;%%
\bibitem{Witten:1998qj}

  E.~Witten,
  ``Anti-de~Sitter Space and Holography,''
{\slshape   Adv.\ Theor.\ Math.\ Phys.\  }{\bf 2} (1998) 253
  [arXiv:hep-th/9802150].
  
%%CITATION = 00203,2,253;%%
\bibitem{Aharony:2008ug}

  O.~Aharony, O.~Bergman, D.~L.~Jafferis and J.~Maldacena,
  ``${\mathcal{N}}\!=6$ Superconformal Chern-Simons-Matter Theories, M2-Branes and Their   Gravity Duals,''
{\slshape   JHEP }{\bf 0810} (2008) 091
  [arXiv:0806.1218 [hep-th]].
  
%%CITATION = JHEPA,0810,091;%%


\end{thebibliography}
\end{document}